\documentclass{article}

\usepackage{PRIMEarxiv}

\usepackage{float}
\usepackage{hyperref}
\usepackage{xspace}
\usepackage{algorithmicx}
\usepackage{algorithm}
\usepackage[noend]{algpseudocode}
\usepackage{multicol}
\usepackage{mdframed}
\usepackage{todonotes}
\usepackage{enumitem}
\usepackage{cite}
\usepackage{subfigure}

\usepackage{listings}
\usepackage{caption}

\newcommand{\punt}[1]{}
\newcommand{\cmnt}[1]{}
\newcommand{\ignore}[1]{}

\definecolor{darkblue}{rgb}{0.0, 0.0, 0.55}

\newcounter{history}

\newcommand{\op} {operation\xspace}

\newcommand{\cas} {\texttt{\textbf{CAS}}\xspace}
\newcommand{\faa}{\texttt{\textbf{FAA}}\xspace}
\newcommand{\FAA}{fetch-and-increment\xspace}
\newcommand{\CAS}{compare-and-swap\xspace}

\newcommand{\urv} {Uruv\xspace}

\newcommand{\lble} {linearizable\xspace}
\newcommand{\lbty} {linearizability\xspace}

\newcommand{\insertADT}{\textsc{Insert}\xspace}
\newcommand{\remove}{\textsc{Delete}\xspace}
\newcommand{\search}{\textsc{Search}\xspace}
\newcommand{\rangeQuery}{\textsc{RangeQuery}\xspace}

\newcommand \spnote[1] {\todo[inline,size=\footnotesize,color=yellow!20]{Sathya: #1}}

\author{
  Gaurav Bhardwaj \\
  Indian Institute of Technology \\
  Hyderabad.\\
  \texttt{cs19resch11003@iith.ac.in} \\
  \And
  Abhay Jain \\
  Indian Institute of Technology \\
  Hyderabad.\\
  \texttt{jain.abhay666@gmail.com} \\
  \And
  Bapi Chatterjee \\
  Indraprastha Institute of Information Technology\\
  Delhi.\\
  \texttt{bapi@iiitd.ac.in} \\
  \And
  Sathya Peri \\
  Indian Institute of Technology\\
  Hyderabad.\\
  \texttt{sathya\_p@cse.iith.ac.in} 
} 

\begin{document}
\title{Wait-Free Updates and Range Search using Uruv} 
\maketitle
\begin{abstract}
CRUD operations, along with range queries make a highly useful abstract data type (ADT), employed by many dynamic analytics tasks. Despite its wide applications, to our knowledge, no fully wait-free data structure is known to support this ADT. In this paper, we introduce Uruv, a proactive \lble and practical wait-free concurrent data structure that implements the ADT mentioned above. Structurally, Uruv installs a balanced search index on the nodes of a linked list. Uruv is the first wait-free and proactive solution for concurrent B$^+$tree. Experiments show that Uruv significantly outperforms previously proposed lock-free B$^+$trees for dictionary operations and a recently proposed lock-free method to implement the ADT mentioned above. 
\end{abstract}

\section{Introduction}
\label{sec:Introduction}
With the growing size of main memory, the in-memory big-data analytics engines are becoming increasingly popular \cite{zhang2015memory}. Often the analytics tasks are based on retrieving keys from a dataset specified by a given range. Additionally, such applications are deployed in a streaming setting, e.g., Flurry \cite{flurry}, where the dataset ingests real-time updates. Ensuring progress to every update would be attractive for many applications in this setting, such as financial analytics \cite{tian2015latency}. The demand for real-time high-valued analytics, the powerful multicore CPUs, and the availability of large main memory together motivate designing scalable concurrent data structures to utilize parallel resources efficiently.

%

It is an ever desirable goal to achieve \textit{maximum progress} of the concurrent operations on a data structure. The maximum progress guarantee -- called \textit{wait-freedom} \cite{Herlihy+:OnNatProg:opodis:2011} -- ensures that each concurrent non-faulty thread completes its operation in a finite number of steps. Traditionally, wait-freedom has been known for its high implementation cost and subsided performance. Concomitantly, a weaker guarantee that some non-faulty threads will finitely complete their operations -- known as \textit{lock-freedom} -- has been a more popular approach. However, it has been found that the lock-free data structures can be transformed to \textit{practical} wait-free \cite{kogan2012methodology} ones with some additional implementation and performance overhead. Progress promises of wait-free data structures make their development imperative, to which a practical approach is to co-design them with their efficient lock-free counterpart. While a progress guarantee is desirable, consistency of concurrent operations is a necessity. The most popular consistency framework is \textit{\lbty} \cite{herlihy1990linearizability}, i.e., every concurrent operation emerges taking effect at an atomic step between its invocation and return.  

In the existing literature, the lock-free data structures such as k-ary search trees \cite{brown2012range}, and the lock-based key-value map KiWi \cite{basin2017kiwi} provide range search. In addition, several generic methods of concurrent range search have been proposed. Chatterjee \cite{chatterjee2017lock} presented a lock-free range search algorithm for lock-free linked-lists, skip-lists, and binary search trees. Arbel-Raviv and Brown \cite{arbel2018harnessing} proposed a more generic approach associated with memory reclamation that fits into different concurrency paradigms, including lock-based and software transactional memory-based data structures. Recently, two more approaches -- bundled-reference \cite{nelson2021bundled} and constant time snapshots \cite{wei+:PPOPP:2021} -- were proposed along the same lines of generic design. Both these works derive from similar ideas of expanding the data structure with versioned updates to ensure \lbty of scans. While the former stores pointers with time-stamped updates, the latter adds objects to nodes time-stamped by every new range search. Moreover, bundled-reference~\cite{nelson2021bundled} design requires locks in every node.
 
In most cases, for example \cite{basin2017kiwi, chatterjee2017lock, nelson2021bundled, wei+:PPOPP:2021}, the range scans are unobstructed even if a concurrent modification (addition, deletion, or update) to the data structure starves to take even the first atomic step over a shared node or pointer. A reader would perceive, indeed for good reasons, that once the modifications are made wait-free the entire data structure will become wait-free. However, to our knowledge, none of these works actually investigates how trivial or non-trivial it would be to arrive at the final implementation of concurrent wait-free CRUD and range-search. This is exactly where our work contributes.

\vspace{2mm}
\noindent\textbf{\large Proposed wait-free \lble proactive data structure}
\vspace{2mm}

In principle, \urv's design derives from that of a B$^+$Tree \cite{comer+:ACM1979}, a self-balancing data structure. However, we need to make the following considerations:

\noindent
\textbf{Wait-freedom:} Firstly, to ensure wait-freedom to an operation that needs to perform at least one \cas execution on a shared-memory word, it must \textit{announce} its invocation \cite{kogan2012methodology}. Even if delayed, the announcement has to happen on realizing that the first \cas was attempted a sufficient number of times, and yet it starved \cite{kogan2012methodology}. The announcement of invocation is then followed by a \textit{guaranteed help} by a concurrent operation at some finite point \cite{kogan2012methodology}. 

\noindent
\textbf{Linearizability:} Now, to ensure \lbty of a scan requires that its output reflects the relevant changes made by every update during its lifetime. The technique of repeated multi-scan followed by validation \cite{brown2012range}, and 
%
  collecting the updates at an augmented object, such as RangeCollector in \cite{chatterjee2017lock}, to let the range search incorporate them before it returns, have been found scaling poorly \cite{brown2012range,chatterjee2017lock}.
   Differently, multi-versioning of objects, for example \cite{nelson2021bundled}, can have a (theoretical) possibility to stockpile an infinite number of versioned pointers between two nodes.   
Interestingly, \cite{arbel2018harnessing} exploits the memory reclamation mechanism to synchronize the range scans with delete operations via logically deleted nodes. However, for lock-freedom, they use a \textit{composite primitive} double-compare-single-swap (DCSS). In comparison, \cite{wei+:PPOPP:2021} uses only single-word \cas. However, managing the announcement by a starving updater that performs the first \cas to introduce a versioned node to the data structure requires care for a wait-free design.

\noindent
\textbf{Node Structure:} The ``fat" (array-based) data nodes, for example Kiwi \cite{basin2017kiwi}, improve traversal performance by memory contiguity \cite{kowalski2020high}. 
However, the benchmarks in \cite{wei+:PPOPP:2021} indicate that it does not necessarily help as the number of concurrent updates picks up. 
Similarly, the lock-free B+trees by Braginsky and Petrank \cite{braginsky+:SPAA:2012} used memory chunks, and our experiments show that their method substantially underperforms. Notwithstanding, it is wise to exploit memory contiguity wherever there could be a scope of ``slow" updates in a concurrent setting.

\noindent
\textbf{Proactive maintenance:} Finally, if the number of keys in a node exceeds (falls short of) its maximum (minimum) threshold after an insertion (deletion), it requires splitting (merging). The operation splitting the node divides it into two while adding a key to its parent node. It is possible that the split can percolate to the root of the data structure if the successive parent nodes reach their respective thresholds. Similarly, merging children nodes can cause cascading merges of successive parent nodes. With concurrency, it becomes extremely costly to tackle such cascaded split or merge of nodes from a leaf to the root. An alternative to this is a \textit{proactive approach} which checks threshold of nodes whiles traversing down a tree every time; if a node is found to have reached its threshold, without waiting for its children, a pre-emptive split or merge is performed. As a result, a restructure remains localized. To our knowledge, no existing concurrent tree structure employs this proactive strategy.

With these considerations, we introduce a key-value store \textbf{Uruv}
(or, Uruvriksha) for wait-free updates and range search. More specifically, 
\begin{enumerate}[label=(\alph*)]
	\item \urv stores keys with associated values in leaf nodes structured as linked-list. The interconnected leaf nodes are indexed by a balanced tree of fat nodes, essentially, a classical B+ Tree \cite{comer+:ACM1979}, to facilitate fast key queries. (Section \ref{sec:Preliminaries})
	\item The key-nodes are augmented with list of versioned nodes to facilitate range scans synchronize with updates. (Section \ref{sec:MainAlgo})
	\item \urv uses single-word \cas primitives. Following the fast-path-slow-path technique of Kogan and Petrank \cite{kogan2012methodology}, we \textit{optimize} the helping procedure for wait-freedom. (Section  \ref{sec:WaitFree}). We prove \lbty and wait-freedom and present the upper bound of step complexity of operations. (Section \ref{sec:Correctness})
	\item Our C++ implementation of \urv significantly outperforms existing similar approaches -- lock-free B+tree of \cite{braginsky+:SPAA:2012}, and OpenBWTree \cite{wang2018building} for dictionary operations. It also outperforms a recently proposed method by Wei et al. \cite{wei+:PPOPP:2021} for concurrent workloads involving range search. (Section \ref{sec:Results})  
\end{enumerate}

\section{Preliminaries}
\label{sec:Preliminaries}
We consider the standard shared-memory model with atomic \texttt{read}, \texttt{write}, \texttt{\faa} (\FAA),  and \texttt{\cas} (\CAS) instructions. \urv implements a key-value store $(\mathcal{K},\mathcal{V})$ of keys $K\in\mathcal{K}$ and their associated values $V\in\mathcal{V}$.

\textbf{The Abstract Data Type (ADT):} We consider an ADT $\mathcal{A}$ as a set of operations: $\mathcal{A}$ = $\{\insertADT(K,V)$, ~$\remove(K)$, ~$\search(K)$,  ~$\rangeQuery(K1,K2)\}$
\begin{enumerate}[topsep=0pt,itemsep=0pt,parsep=0pt,partopsep=0pt]
\item An $\insertADT(K, V)$ inserts the key $K$ and an associated value $V$ if $K\notin\mathcal{K}$.
\item A $\remove(K)$ deletes the key $K$ and its associated value if $K\in\mathcal{K}$.
\item A $\search(K)$ returns the associated value of key $K$ if $K\in\mathcal{K}$; otherwise, it returns $-1$. It does not modify $(\mathcal{K},\mathcal{V})$.
\item A $\rangeQuery(K_1,K_2)$ returns keys $\{K\in\mathcal{K}:K_1{\le}K{\le}K_2\}$, and associated values without modifying $(\mathcal{K},\mathcal{V})$; if no such key exists, it returns $-1$.
\end{enumerate}
\subsection{Basics of Uruv's Lock-free Linearizable Design}
\label{subsec:Design}
\begin{figure}
\begin{minipage}[t]{0.60\linewidth}  
    \centering
    \includegraphics[width = 0.99\textwidth]{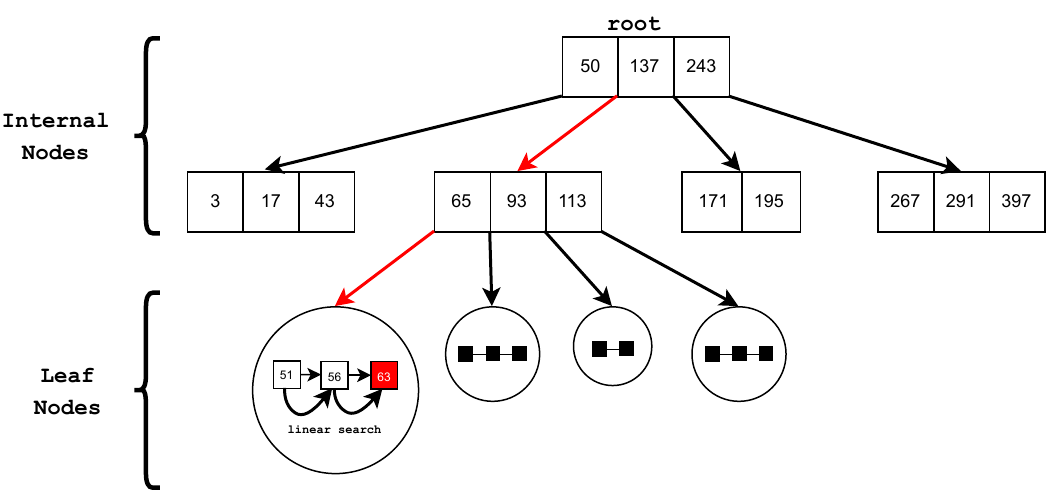}
    \caption{\scriptsize Example of Uruv's design. In this example, a search operation is being performed wherein the red arrows indicate a traversal down Uruv, and we find the key, highlighted red, in the linked-list via a linear search.}
    \label{fig:tree-design}
\end{minipage}
\begin{minipage}[t]{0.39\linewidth}
    \centering
    \includegraphics[width = 0.95\textwidth]{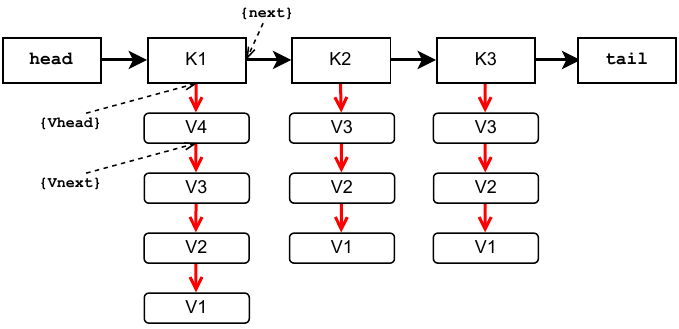}
    \caption{Versioned key nodes}
    \label{fig:vers_list}
\end{minipage}
\end{figure}

\urv derives from a B$^+$Tree \cite{comer+:ACM1979}, a self-balancing data structure. However, to support \lble range search operations, they are equipped with additional components. 
The key-value pairs in \urv are stored in the \textit{key nodes}. A \textit{leaf node} of \urv is a sorted linked-list of key nodes. Thus, the leaf nodes of \urv differ from the array-based leaf nodes of a B$^+$Tree. The \textit{internal nodes} are implemented by arrays containing ordered set of keys and pointers to its descendant children, which facilitate traversal from the root to key nodes. A search path in \urv is shown in Figure \ref{fig:tree-design}.

\section{Lock-Free Algorithm}
\label{sec:MainAlgo}

\subsection{The structures of the component nodes}
Here we first describe the structure of the nodes in \urv. See Figure \ref{fig:vlfll}. A versioning node is implemented by the objects of type Vnode. A key node as described in the last section, is implemented by the objects of the class llNode. Nodes of type llNode make the linked-list of a leaf-node which is implemented by the class VLF\_LL.

The leaf and internal nodes of \urv inherit the Node class. See Figure \ref{fig:tree-class}. An object of class Node of \urv, hereafter referred to as a node object, keeps count of the number of keys. A node object also stores a boolean to indicate if it is a leaf node. A boolean variable `frozen' helps with ``freezing a node'' while undergoing a split or merge in a lock-free updatable setting. A thread on finding that a node is frozen, helps the operation that triggered the freezing.

\begin{figure}[h]
    \centering
    \begin{minipage}[t]{0.32\linewidth}
    \begin{lstlisting}[basicstyle=\scriptsize]
Vnode{
    value_t value;
    int ts;
    Vnode* nextv;
}
    \end{lstlisting}
    \end{minipage}
    \begin{minipage}[t]{0.32\linewidth}
    \begin{lstlisting}[basicstyle=\scriptsize]
llNode{
      key_t key;
      Vnode* vhead;
      llNode* next;
}
    \end{lstlisting}
    \end{minipage}
     \begin{minipage}[t]{0.32\linewidth}
    \begin{lstlisting}[basicstyle=\scriptsize]
VLF_LL{

      llNode* head;
    
}
    \end{lstlisting}
    \end{minipage}
\caption{Versioned Lock-Free Linked-List Data Structure}
\label{fig:vlfll}
\end{figure}

Every leaf node has three pointers $next$, $newNext$ and a pointer to version list $ver\_head$ and one variable $ts$ for the timestamp. The $next$ pointer points to the next adjacent leaf node in \urv. When a leaf node is split or merged, the $newNext$ pointer ensures leaf connection. A new leaf node is created to replace it when a leaf node is balanced. Using the $newNext$ pointer, we connect the old and new leaf nodes. When traversing the leaf nodes for \rangeQuery with $newNext$ set, we follow $newNext$ instead of $next$ since that node has been replaced by a newer node, ensuring correct traversal. The initial $ts$ value is associated with the construction of the leaf node. 

\begin{figure}
	\centering
	\input{figs/tree_class.tex}
	\caption{The details of object structures}
	\label{fig:tree-class}
\end{figure}

\subsection{Versioned Linked-List}

The description of lock-free \lble implementation of the ADT operations \rangeQuery, \insertADT, and \remove requires detailing the versioned linked list. A versioned list holds the values associated with the key held at various periods. Each versioned node (\texttt{Vnode}) in the versioned list has a value, the time when the value was modified, and a link to the previous version of that key. Versioned linked-list information may be seen in Figure \ref{fig:vers_list} and Figure \ref{fig:vlfll}. The versioned list's nodes are ordered in descending order by the time they have been updated. Compared to the \cite{harris2001pragmatic}, there is no actual delinking of nodes; instead, we utilise a tombstone value (a special value not associated with any key) to indicate a deleted node. Moreover, deleting a node requires no help since there is no delinking. Although, for memory reclamation, we retain a record of active \rangeQuery and release nodes that are no longer needed. Any modification to the versioned linked list atomically adds a version node to the \texttt{vhead} of \texttt{llNode} using \texttt{CAS}. 

\subsection{Traversal and Proactive maintenance in \urv}

\begin{figure}
\begin{minipage}{.32\linewidth}
    \centering
    \includegraphics[width = 1 \linewidth]{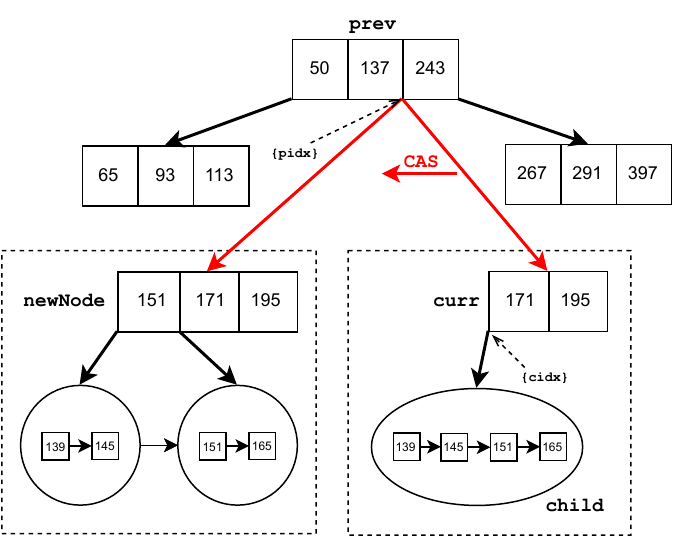}
    \textbf{(a)}
\end{minipage}
\begin{minipage}{.32\linewidth}
    \centering
    \includegraphics[width = 1 \linewidth]{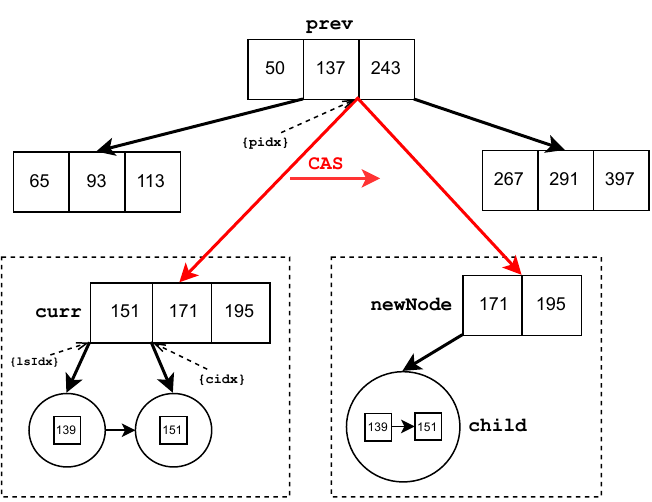}
    \textbf{(b)}
\end{minipage}
\begin{minipage}{.32\linewidth}
      \centering
      \includegraphics[width = 1 \linewidth]{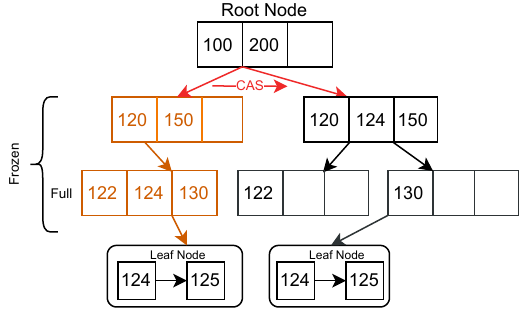}
    \textbf{(c)}
\end{minipage}
\caption{(a) Split Leaf, (b) Merge Leaf, (c) Split Internal}
\label{fig:Tree Operations}
\end{figure}

We traverse from root to a leaf following the order provided by the keys in the internal nodes. In each internal node a binary search is performed to determine the appropriate child pointer. While traversal in 
\insertADT and \remove operations, we follow the proactive approach  as described earlier. Essentially, if we notice that a node's key count has violated the maximum/minimum threshold, we instantly conduct a split/merge action, and the traversal is restarted. The proactive maintenance is shown in  Figure\ref{fig:Tree Operations}.

\subsection{ADT Operations}
\begin{figure}[H]

\algrenewcommand\alglinenumber[1]{\scriptsize #1:}
\begin{multicols}{2}
\begin{algorithmic}[1]
\renewcommand{\algorithmicprocedure}{}
\label{Read}
\scriptsize
\Procedure{\textbf{Insert}}{key, value}
\State{\texttt{retry}:}
\State{$\textup{Node* } curr := root$}
\If{$curr =$ \textup{\texttt{nullptr}}}
    \State{$\textup{Node* }\ nLeaf\rightarrow$\texttt{insLeaf}($key$, $value$)}
    \If{\textup{!}$\,root$.\textup{\texttt{CAS}}($curr, nLeaf$)}
        \label{alg1:line5}
        \State{\textup{\textbf{goto} \texttt{retry}}}
        \label{alg1:line6}
    \Else { \textup{\textbf{return}}}
    \EndIf
\EndIf
\State{$curr :=$ \textup{\texttt{balanceRoot(curr)}}}
\label{alg1:line8}
\If{{\textup{!}}$\,curr$}{\textup{ \textbf{goto} \texttt{retry}}}
\EndIf
\State{$\textup{Node* } prev, child :=$ \texttt{nullptr}}
\State{int $pidx, cidx$}
\While{\textup{!}$curr \rightarrow isLeaf$}
    \label{alg1:line12}
    \If{$curr \rightarrow helpIdx \neq -1$}
        \label{alg1:line13}
        \State{$\textup{Node*} \, res:=$ \texttt{help}($prev, pidx, curr$)}
        \If{$res$} {$curr := res$}
        \Else { \textup{\textbf{goto} \texttt{retry}}}
        \EndIf
    \EndIf
    \label{alg1:line16}
    \State{{$cidx$ \textup{is set to the index of appropriate child based on} $key$ using Binary Search}}
    \label{alg1:line17}
    \State{$child := curr \rightarrow ptr[cidx]$}
    \If{$child\rightarrow isLeaf  \,\&\&\ child\rightarrow frozen$}
        \State{$\,curr \rightarrow$ \texttt{freezeInternal()}}
        \label{alg1:line20}
        \If{\textup{!}$\,curr \rightarrow$\textup{\texttt{setHelpIdx$(cidx)$}}} 
        \State{\textup{\textbf{goto} \texttt{retry}}}
        
        \EndIf
        \State{$\textup{Node*} newNode:=child\rightarrow$ \texttt{balanceLeaf}($prev, pidx, curr, cidx)$}
        \label{alg1:line23}
        \If{$newNode$} \State{\textbf{then} $curr := newNode$}
        \label{alg1:line24}
        \Else{ \textup{\textbf{goto}        \texttt{retry}}}
        \label{alg1:line25}
        \EndIf
       
    \ElsIf{\textup{!}$\,child \rightarrow isLeaf \&\&\ child \rightarrow count \geq MAX$}
    {
        \label{alg1:line26}
        \State{$\,curr \rightarrow$ \texttt{freezeInternal()}}
        \label{alg1:line27}
        \If{\textup{!}$\,curr\rightarrow$\textup{\texttt{setHelpIdx$(cidx)$}}} {
         \State{\textup{\textbf{goto} \texttt{retry}}}
            \label{alg1:line29}
        }
        \EndIf
        \State{$\textup{Node* } newNode := child \rightarrow$ \texttt{splitInternal}($prev, pidx, curr, cidx)$}
        \label{alg1:line30}
        \If{$newNode$} \State{\textbf{then} $curr := newNode$}
        \label{alg1:line24}
        \Else{ \textup{\textbf{goto}        \texttt{retry}}}
        \label{alg1:line25}
        \EndIf
    }
    \EndIf
     
    \State{$prev := curr$}
    \State{$curr := child$}
    \label{alg1:line34}
    \State{$pidx := cidx$}

\EndWhile

\State{$res:=$ $curr \rightarrow$ \textup{\texttt{insertLeaf}}($key, value$)}
\label{alg1:insert_tree}
\If{$res = $ \textup{Failed}}
    \State{\textup{\textbf{goto}        \texttt{retry}}}
\Else{ \textbf{return} \textup{res}}
\EndIf
%
\EndProcedure
\algstore{insert}
\end{algorithmic}
\end{multicols}
\caption{Pseudocode of \insertADT operation}
\label{alg:insert}
\end{figure}

\textbf{An \insertADT operation} starts with performing a traversal as described above to locate the leaf node to insert a key and its associated value. It begins with the root node; if it does not exist, it builds a new leaf node and makes it the root with a \textbf{\texttt{CAS}}. If it cannot update the root, another thread has already changed it, and it retries insertion. Method \textup{balanceRoot} splits the root if needed and replaces it with a new root using \texttt{CAS}. If \texttt{CAS} fails, then some other thread must have changed the root, and it returns null. If there is no need to split the root, it will return the current root. 

Lines \ref{alg1:line13}-\ref{alg1:line16} describe the helping mechanism, which makes the data structure lock-free. If any node helpIdx is set to a value other than -1, then the child node at helpIdx is undergoing the split/merge process. In that case, it will help that child finish its split/merge operation. Method \texttt{help} helps $child$ node in split/merge operation and returns the new $curr$ node if it successfully replaces it using $CAS$; otherwise, it returns null. Then, it performs a binary search over $curr$’s keys at line \ref{alg1:line17} to find the correct child pointer. It copies the child pointer into $child$ and stores its index in the pointer array as $cidx$. 

If the child node is a frozen leaf node or an internal node that has reached the threshold, it performs a split/merge operation. It starts by freezing its parent, $curr$, at line \ref{alg1:line20} by setting a special freezing marker on every child pointer, so that no other thread can change the parent node and cause inconsistency. After freezing the parent, it stores the index of the child pointer in helpIdx of the parent node using $CAS$ so that other threads can help in split/merge operation. If \textup{setHelpIdx} fails, that means some other thread has already set the helpIdx, and it retries. 

\textbf{Restructuring} a $child$ is performed at line \ref{alg1:line23} and \ref{alg1:line30} using \textup{balanceLeaf} and \textup{splitInternal} respectively. \textup{balanceLeaf} performs the split/merge operation on the leaf node based on the number of elements and returns the node replaced by the parent node using $CAS$. Similarly, \textup{splitInternal} splits the internal node and returns the new parent node. If in any of the above methods, $CAS$ is failed, then some other thread must have replaced it, and it will return nullptr and retries at line \ref{alg1:line29} and \ref{alg1:line25}. It repeats the same process until it reaches the leaf node. Once it reaches the leaf node, it performs the insert operation in the leaf node at line \ref{alg1:insert_tree}. It returns on success, otherwise it retries.

\textbf{Insert into a leaf.} In the leaf node, all the updates occur concurrently in the versioned linked list. It first checks if the leaf node is  frozen. If it is, it returns \textup{"Failed"}, realizing that another thread is trying to balance this node. If the node’s count has reached the maximum threshold, it freezes it and returns \textup{"Failed"}. Leaf node is frozen by setting a special freezing mark on \textup{llnode} \textit{next} pointer and the \textit{vhead} pointer. In both the cases, when it returns \textup{"Failed"} insertion will be retried after balancing it. Otherwise, it would insert the key into the versioned lock-free linked list. If another thread is concurrently freezing the leaf node, the insertion into the linked list might fail. If it fails, it will again return \textup{"Failed"} and retries the insertion. If the key is already present in the linked list, it updates that key’s version by adding a new version node in the version list head with a new value. Else it will create a new node in the linked list containing the key and its value in the version node. After the key is inserted/updated in the linked list, its timestamp is set to the current timestamp, which is the linearization point for insertion in the tree.

\textbf{A \remove \op} follows a similar approach as \insertADT. It traverses the tree to the leaf node, where the key is present. The difference in traversal with respect to \insertADT \op is that at line \ref{alg1:line26}, instead of checking the max threshold, it checks for the minimum threshold. Instead of splitting the internal node  at line \ref{alg1:line30}, it merges the internal node. Once a leaf node is found, it checks whether the key is in the linked list. If it is in the linked list, it will update a tombstone value in the version list to mark that key as deleted. If the key is absent, it returns \textup{"Key not Present"}.

\textbf{Delete from Leaf.} If the key is present, this \op creates a versioned node with a tombstone value to set it as deleted. Just like inserting the new versioned node its timestamp is set to the current timestamp. If the key is not present in the linked list it simply returns "Key Not Present". Detail pseudocode is deferred to Appendix A.

\textbf{\search \op.} Traversal to a leaf node in case of searching doesn't need to perform any balancing. After finding the leaf node, it checks the key in the linked list; if it is present, it returns the value from the version node from the head of the list; otherwise, it simply returns \textup{"Key not Present"}. Before reading the value from the versioned node it checks if the timestamp is set or not. If it is not set, it sets the timestamp as the current timestamp before reading the value. 

\textbf{\rangeQuery.} A range query returns keys and their associated values by a given range from the data. \urv supports a linearizable range query employing a multi-version linked list augmented to the nodes containing keys. This approach draws from Wei et al.\cite{wei+:PPOPP:2021}’s work.  
A global timestamp is read and updated every time a range query is run. The leaf node having a key larger than or equal to the beginning of the supplied range is searched after reading the current time. Then, it chooses a value for the relevant key from the versioned list of values. Figure \ref{fig:Tree Operations}(c) depicts a versioned linked list, with the higher versions representing the most recent modifications.

By iterating over each versioned node individually, it selects the first value in the list whose timestamp is smaller than the current one. This means that the value was changed before the start of the range query, making it consistent. It continues to add all keys and values that are less than or equal to the end of the given range. Because all of the leaf nodes are connected, traversing them is quick. After gathering the relevant keys and values, the range query will produce the result. 

As a leaf node could be under split or merge, for every leaf node that we traverse, we first check whether their $newNext$ is set. If it is and the leaf pointed to by $newNext$ has a timestamp lower than the range query’s timestamp, it traverses the $newNext$ pointer. This ensures that our range query collects data from the correct leaf nodes. Were the timestamp not part of the leaf node, there is a chance that the range query traverses $newNext$ pointers indefinitely due to repeated balancing of the leaf nodes.
\section{Wait-Free Construction}
\label{sec:WaitFree}
We now discuss a wait-free extension to the presented lock-free algorithm above. Wait-freedom is achieved using fast-path-slow-path method \cite{Timnat+:WFLis:opodis:2012}. More specifically, a wait-free operation starts exactly as the lock-free algorithm. 
This is termed as the \textit{fast path}. 
If a thread cannot complete its operation even after several attempts, it enters the \textit{slow path} by announcing that it would need help.
 To that effect, we maintain a global \texttt{stateArray} to keep track of the operations that every thread currently needs help with. 
 In the slow path, an operation first publishes a \texttt{State} object containing all the information required to help complete its operation.

For every thread that announces its entry to the slow path, it needs to find helpers. After completing some fixed number of fast path operations, every thread will check if another thread needs some help. This is done by keeping track of the thread to be helped in a thread-local \texttt{HelpRecord} object presented in Figure \ref{fig:WF_DS}. After completing the $nextCheck$ amount of fast path operations, it will assist the $currTid$. Before helping, it checks if $lastPhase$ equals $phase$ in $currTid$'s \texttt{stateArray} entry. If it does, the fast path thread will help execute the wait-free implementation of that operation; otherwise, $currTid$ doesn't require helping as its entry in the \texttt{stateArray} has changed, meaning the operation has already been completed. In the worst case, if the helping thread also faces massive contention, every available thread will eventually execute the same operation, ensuring its success.

\begin{figure}[ht]
    \centering
    \input{figs/WFstructure}
    \caption{Data structures used in wait-free helping}
    \label{fig:WF_DS}
\end{figure}

Notice that when data and updates are uniformly distributed, the contention among threads is low, often none. Concomitantly, in such cases, a slow path by any thread is minimally taken. 

\textbf{Wait-free \insertADT.} Traversal in Wait-free \insertADT is the same as that in the lock-free \insertADT as mentioned in Section \ref{sec:MainAlgo}. While traversal a thread could fail the \textbf{\texttt{CAS}} operation in a split/merge operation of a node and would need to restart traversal from the root again. At first glance, this would appear to repeat indefinitely, contradicting wait-freedom, but this operation will eventually finish due to helping. If a thread repeatedly fails to traverse Uruv due to such failure, every other thread will eventually help it find the leaf node. Once we reach the leaf node, we add the key to the versioned linked list as described below. There are two cases - either a node containing the key already exists, or a node does not exist. 

In the former case, we need to update the linked list node’s \textit{vhead} with the versioned node, \textit{vnode}, containing the new value using \texttt{CAS}.
The significant difference between both methods is the usage of a shared \texttt{Vnode} from the \texttt{stateArray} in wait-free versus a thread local \texttt{Vnode} in lock-free. Every thread helping this insert will take this \textit{vnode} from the \texttt{stateArray} and first checks the variable finished if the operation has already finished. They then check if the phase is the same in the \texttt{stateArray}, and \textit{vnode}’s timestamp is set or not. If either is not true, some other thread has already completed the operation, and they mark the operation as finished. Else, they will try to update the \textit{vhead} with \textit{vnode} atomically. After inserting the \textit{vnode}, it initialises the timestamp and sets the finished to be true.

In the latter case, we create a linked list node, \textit{newNode},  and set its \textit{vhead} to the \textit{vnode} in the \texttt{stateArray} entry. It tries adding \textit{newNode} like the lock-free linked list’s insert. If it is  successful, the timestamp of \textit{vnode} is initialized, and the finished is set to true in the \texttt{stateArray}. We have discussed some race conditions in Appendix \ref{race} due to limited space in this section.

\textbf{Wait-free \remove.} \remove \op follows the same approach as \insertADT. If the key is not present in the leaf node, it returns \textup{"Key Not Present"} and sets the finished to be true. Otherwise, it will add the \textit{vnode} from \texttt{stateArray} similar to wait-free \insertADT. The only difference is that the \textit{vnode} contains the tombstone value for a deleted node. Details on this method are deferred to Appendix \ref{sec:wait-free-appendix}.

\ignore{
Like insertion, we find the leaf node possibly containing the key to be deleted. Once the leaf node is found, it searches for the node to be deleted and updates the \texttt{stateArray} atomically at line \ref{alg10:line142}. If the node does not exist, there is nothing to delete, and the operation is completed. Otherwise, the \textit{vnode} in the \texttt{stateArray} replaces the \textit{vhead} of the linked list node via the method, \texttt{wfLLInsert} at line \ref{alg10:line151}, described above. Deleting from Uruv has the advantage of having no physical deletion of a node. Nodes that are not required will only be physically removed while doing a split or merge operation. For more algorithmic details please refer to Algorithm \ref{wfdeletefromleaf} in the appendix.

}

\textbf{\search and \rangeQuery.} Neither operation modifies Uruv nor helps any other operation; hence their working remain as explained in Section \ref{sec:MainAlgo}.

\section{Correctness and Progress Arguments}
\label{sec:Correctness}
To prove the correctness of Uruv, we have shown that Uruv is linearizable by describing \textit{linearization points} (LPs) that map any concurrent setting to a sequential order of said operations. We discuss them in detail below.

\subsection{Linearization Points}

As explained earlier, we traverse down Uruv to the correct leaf node and perform all operations on the linked list in that leaf. Therefore, we discuss the LPs of the versioned linked list.
\vspace{-0.2cm}\paragraph*{\textbf{\insertADT:}} There are two cases. If the key does not exist, we insert the key into the linked list. However, the timestamp of the \texttt{vnode} is not set, so the \texttt{LP} for \insertADT \op is when the timestamp of \texttt{vnode} is set to the current timestamp. This can be executed either just after the insertion of the key in the linked list or by some other thread before reading the value from \texttt{vnode}.

If the key already exists, we update its value by atomically replacing a new versioned node by its current $vhead$. After successfully changing the $vhead$, the node’s timestamp is still not set. It can be set just after adding the new versioned node or by some other thread before reading the value from the newly added versioned node. In both the cases the \texttt{LP} is when the timestamp of the versioned node is set to the current timestamp. 

\vspace{-0.3cm}\paragraph*{\textbf{\remove.}} There are two cases. If the key does not exist, then there is no need to delete the key as it does not exist. Therefore, the \texttt{LP} would be where we last read a node from the linked list. Instead, if the key exists, the \texttt{LP} will be same as \insertADT when we set the timestamp of the versioned node.

\vspace{-0.2cm}\paragraph*{\textbf{\search}.} There are two cases, first if the key doesn't exist in the linked list, the \search \texttt{LP} would be when we first read the node whose key is greater than the key we are searching for in the linked list. Second, if the key is present in the linked list it reads the value in the versioned node at \texttt{vhead}. So the \texttt{LP} is when we atomically reads the value from the versioned node. If a concurrent insert/delete leads to a split/merge operation, then there is a chance that the search will end up at a leaf node that is no longer a part of Uruv. In that case, the search’s \texttt{LP} would have happened before insert/delete’s \texttt{LP}. Search’s \texttt{LP} remains the same as above.

\vspace{-0.2cm}\paragraph*{\textbf{\rangeQuery}.} \rangeQuery method reads the global timestamp and increment it by 1. So the \texttt{LP} for range query would be the atomic read of global timestamp. The range query’s \texttt{LP} will remain the same regardless of any other concurrent operation.
\ignore{
\subsection{Correctness of the Wait-Free Mechanism}

The LPs for the wait-free algorithm have already been discussed above. There are two essential aspects of the wait-free mechanism that must be carefully analyzed. The first is that a helping thread needs to help the intended operation. There might be a case that the entry in the \texttt{stateArray} is updated while the thread is helping, and it may perform an incorrect operation. To avoid this situation, every helping thread checks if the phase has been updated in the \texttt{stateArray} entry or not. If it has changed, then the entry in the \texttt{stateArray} has been replaced by a later entry, and the intended operation was already done.

The second consideration is that only one thread should be able to complete an operation successfully. Otherwise, two threads may perform the same operation due to \textit{race conditions} which can lead to inconsistency. This has already been discussed above, wherein we show that only one thread can complete any operation preventing this problematic condition.
}
\subsection{Complexity Analysis} Let there be $t$ threads in the shared-memory system. Let the number of attempts after an operation decides to go the slow path way be $f$ and the number of operations performed by a thread between every check of helping a slow peer be $s$. Notice that, when a \cas fails, before reattempting the operation, a thread necessarily helps one of its concurrent peers. Thus, every help gets counted as a failed attempt. With this, the maximum number of restarts that any operation would do is $m=\min(f+st, I_C)$, where $I_C$ is the interval contention, i.e., total number of concurrent operations during the lifetime of the operation \cite{afek1999long}. 

Now, let the  maximum threshold of an internal node (essentially, its size) be $B$ and the maximum size of the linked-list in a leaf node, after which it is decided to be split, be $L$. Then, using the standard worst-case complexity analysis of the hierarchical search, if an update or a search operation remains unobstructed, it will finish in $O(L+\log_B(n))$ number of steps, where $n$  is the number of keys in the data structure at its invocation. 

Let $n_{max,op}$ be the maximum size of the dataset contained in Uruv during the lifetime of an operation $op$. Then counting the maximum number of attempts, the worst case complexity of an update operation will be $O(m(L+\log_B(n_{max,op})))$. Because, the search and range queries do not restart, it is easy to see that their worst case step complexity will be $O(L+\log_B(n_{max,op}))$ and $O(L+\log_B(n_{max,op})+k)$, respectively, where $k$ is the size of the output of range search operations.

Clearly, all the above expressions of worst-case complexity are finite. Therefore, each of the operations in Uruv are wait-free. 
\label{sec:Analysis}

\section{Experiments}
\label{sec:Results}

In this section, we benchmark Uruv against  (a) previous lock-free variants of the B$^+$Tree for updates and search operations (to our knowledge, there are no existing wait-free implementations of the B$^+$Tree, and lock-free B$^+$Trees do not implement range search), and (b) the lock-free VCAS-BST of \cite{wei+:PPOPP:2021}, which is the best-performing data structure in their benchmark. The code of the benchmarks is available at \url{https://github.com/gaurav-bhardwaj03/Uruv/}.


\paragraph{Experimental Setup.} We conducted our experiments on a system with an IBM Power9 model 2.3 CPU packing 40 cores with a minimum clock speed of 2.30 GHz and a maximum clock speed of 3.8 GHz. There are four logical threads for each core, and each has a private 32KB L1 data cache and L1 instruction cache. Every pair of cores shares a 512KB L2 cache and a 10MB L3 cache. The system has 240GB RAM and a 2TB hard disk. The machine runs Ubuntu 18.04.6 LTS. We implement Uruv in C++. Our code was compiled using g++ 11.1.0 with -std=c++17 and linked the pthread and atomic libraries. We take the average of the last seven runs out of 10 total runs, pre-warming the cache the first three times. Our average excludes outliers by considering results closest to the median.

\begin{center}
\begin{tabular}{ccc}
    \multicolumn{3}{c}{\includegraphics[width = 0.7\linewidth]{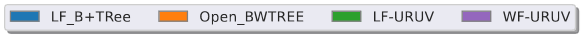}}\\   
     \includegraphics[width=0.33\linewidth]{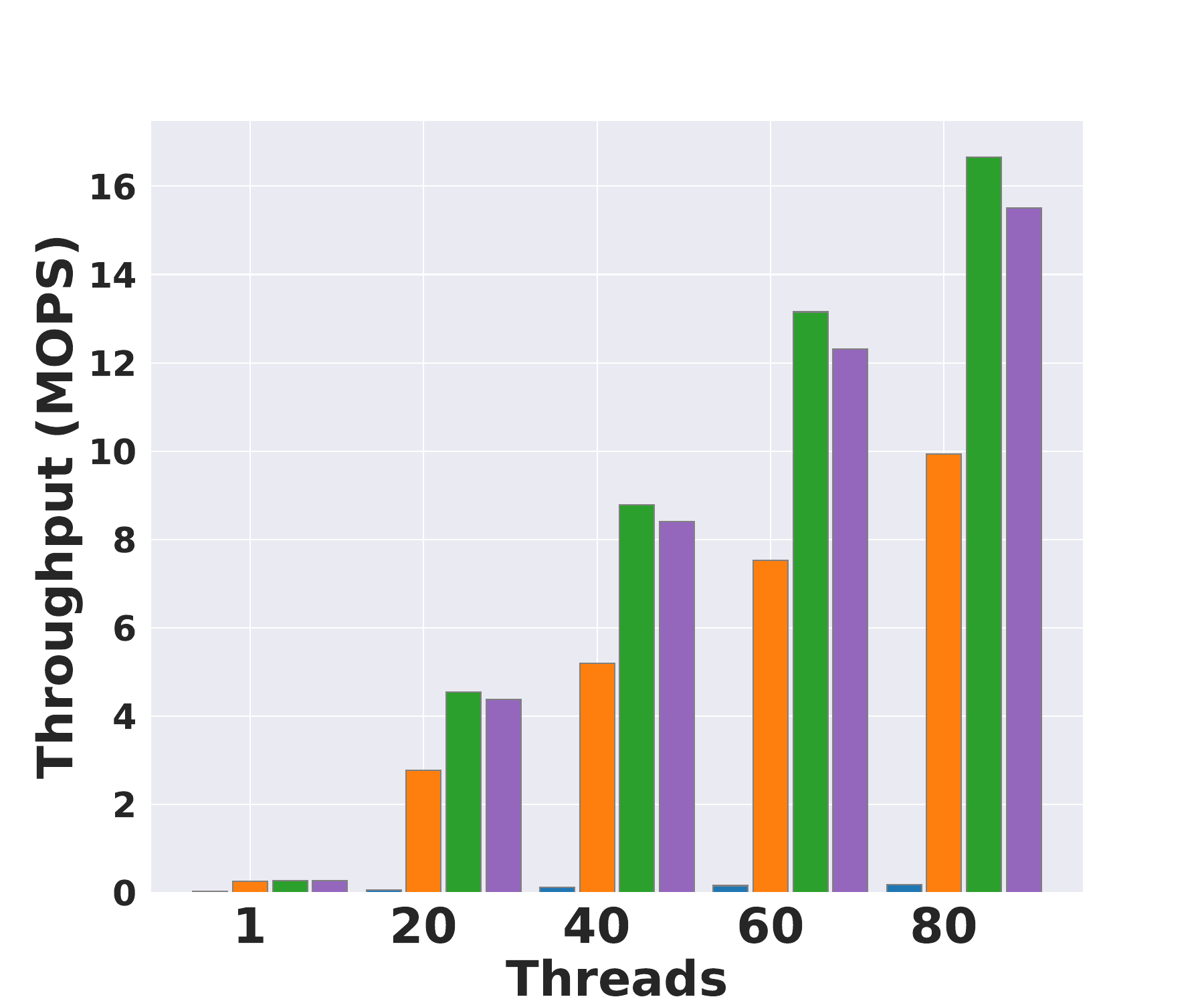} &
     \includegraphics[width=0.33\linewidth]{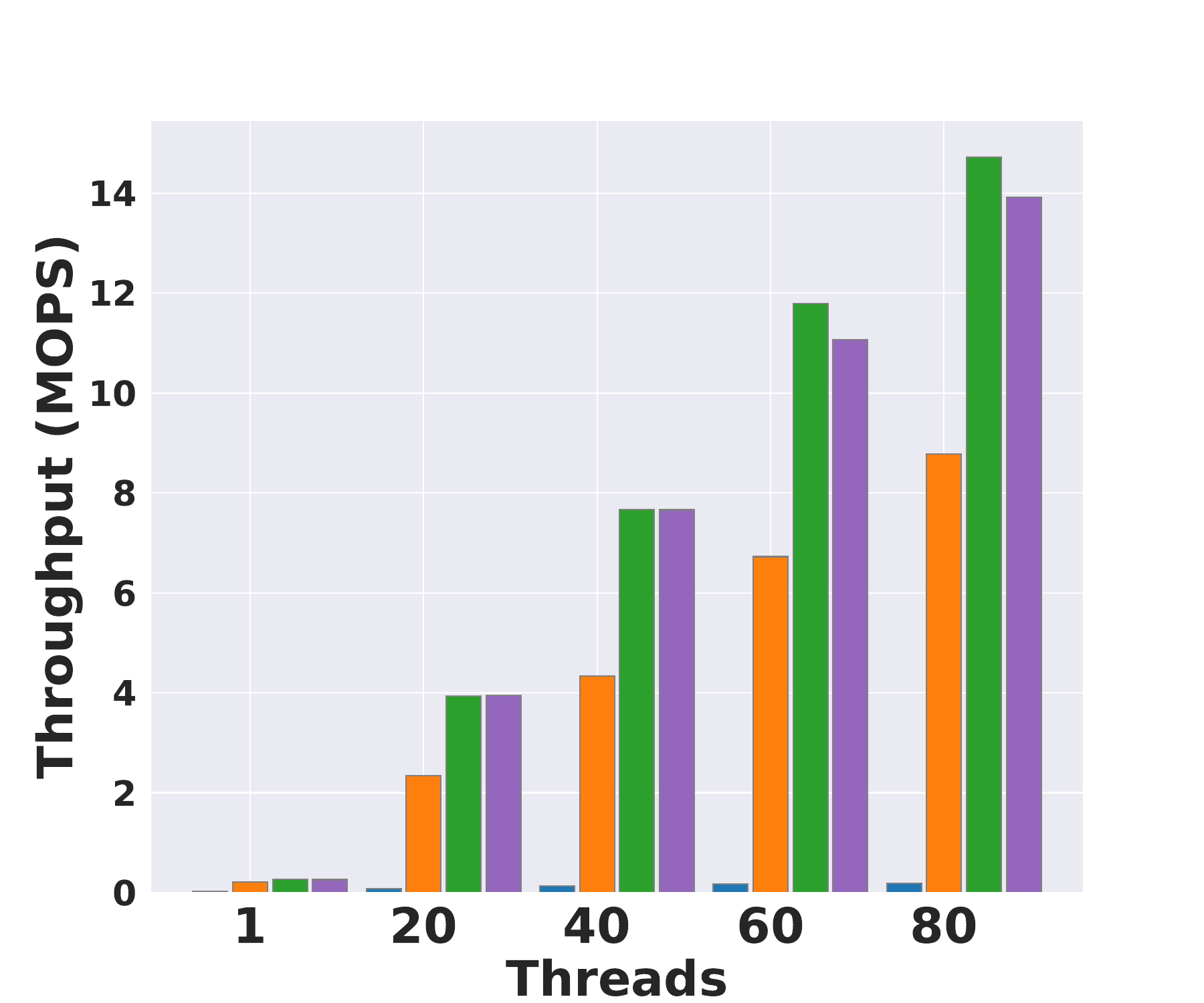} &
     \includegraphics[width=0.33\linewidth]{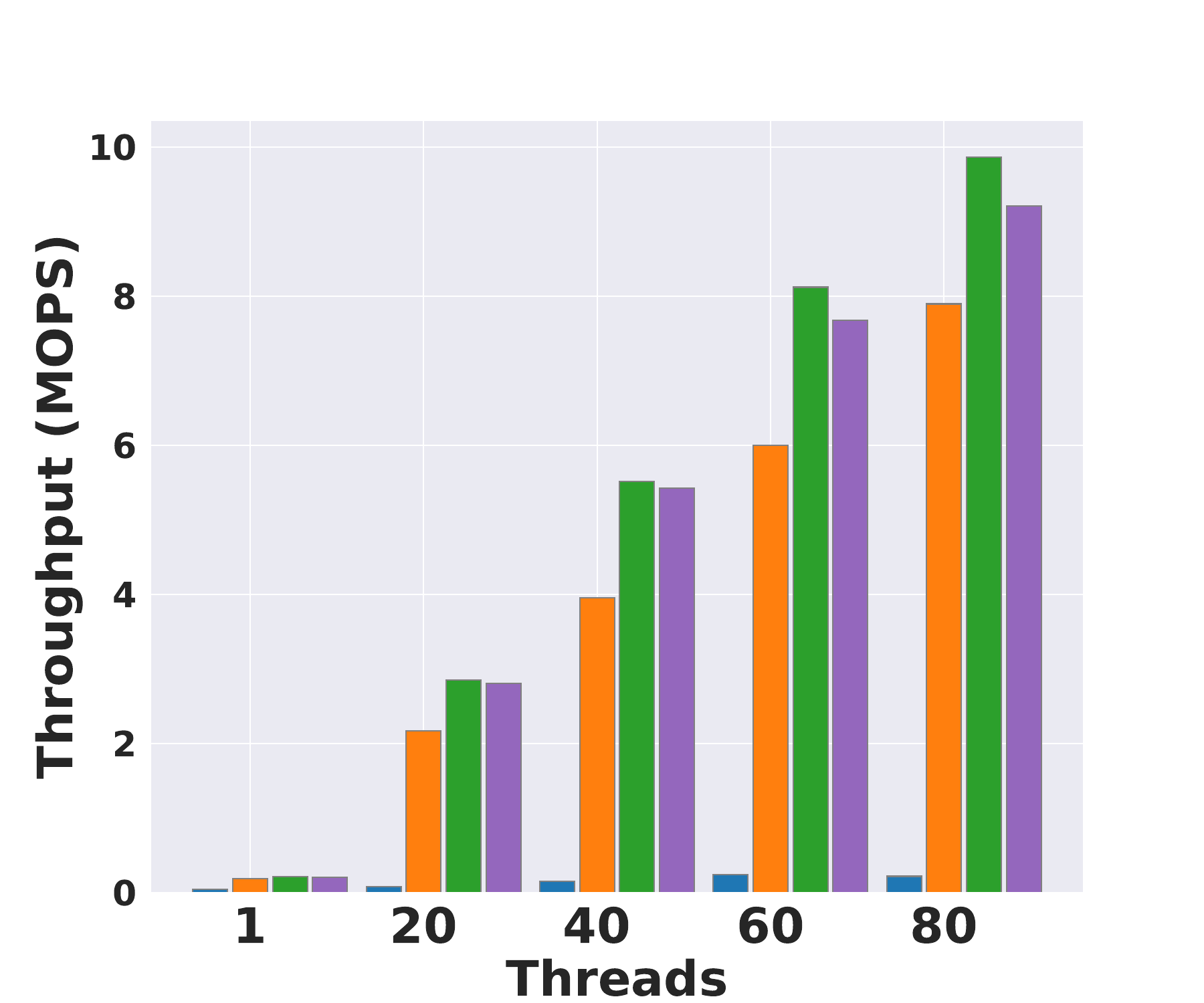}
     \\
     (a) & (b) & (c) \\
\end{tabular}
\scriptsize
\captionof{figure}{\scriptsize The performance of \textbf{Uruv} when compared to \textbf{LF$\_$B+Tree}\cite{braginsky+:SPAA:2012} and \textbf{Open$\_$BwTree}\cite{wang2018building}. Higher is better. The workload distributions are (a) Reads - 100\% (b) Reads - 95\%, Updates - 5\%, and (c) Reads - 50\%, Updates - 50\%}
\label{fig:uruv-anastasia-openbwtree}
\end{center}

\begin{tabular}{ccc}
    \multicolumn{3}{c}{\includegraphics[width=0.7\linewidth]{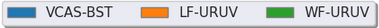}}\\
   \includegraphics[width=0.32\linewidth]{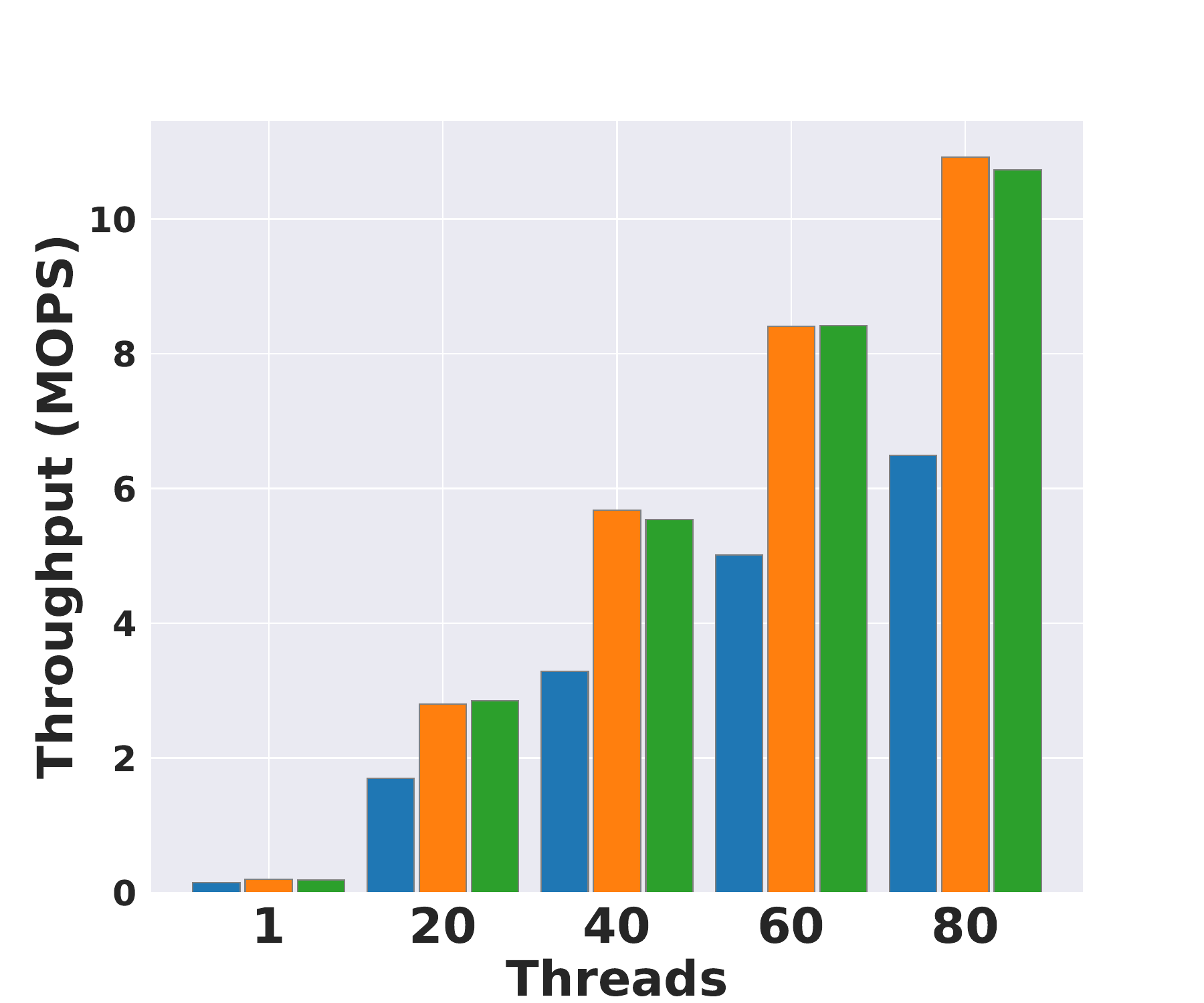}  & 
   \includegraphics[width=0.32\linewidth]{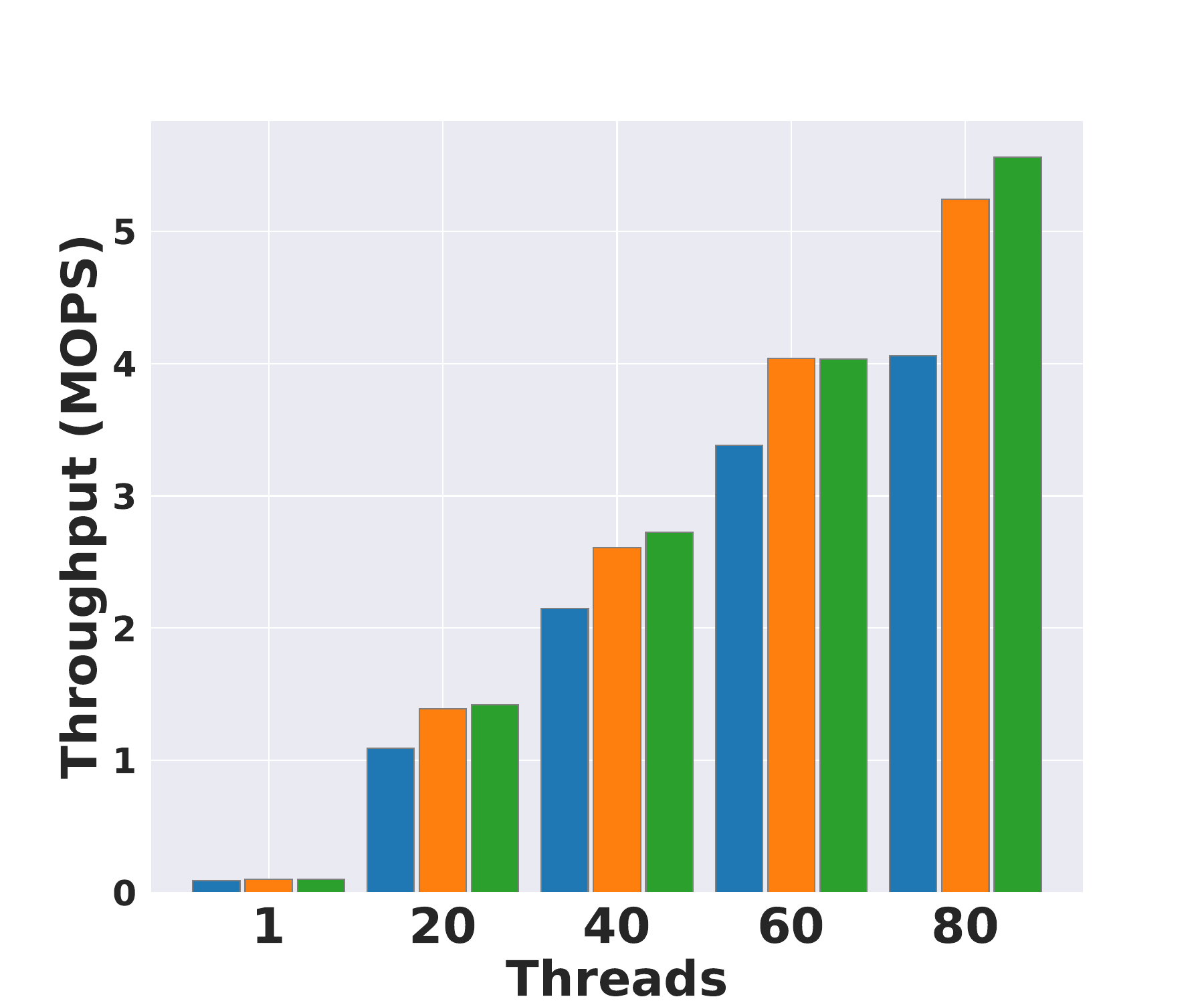} & 
   \includegraphics[width=0.32\linewidth]{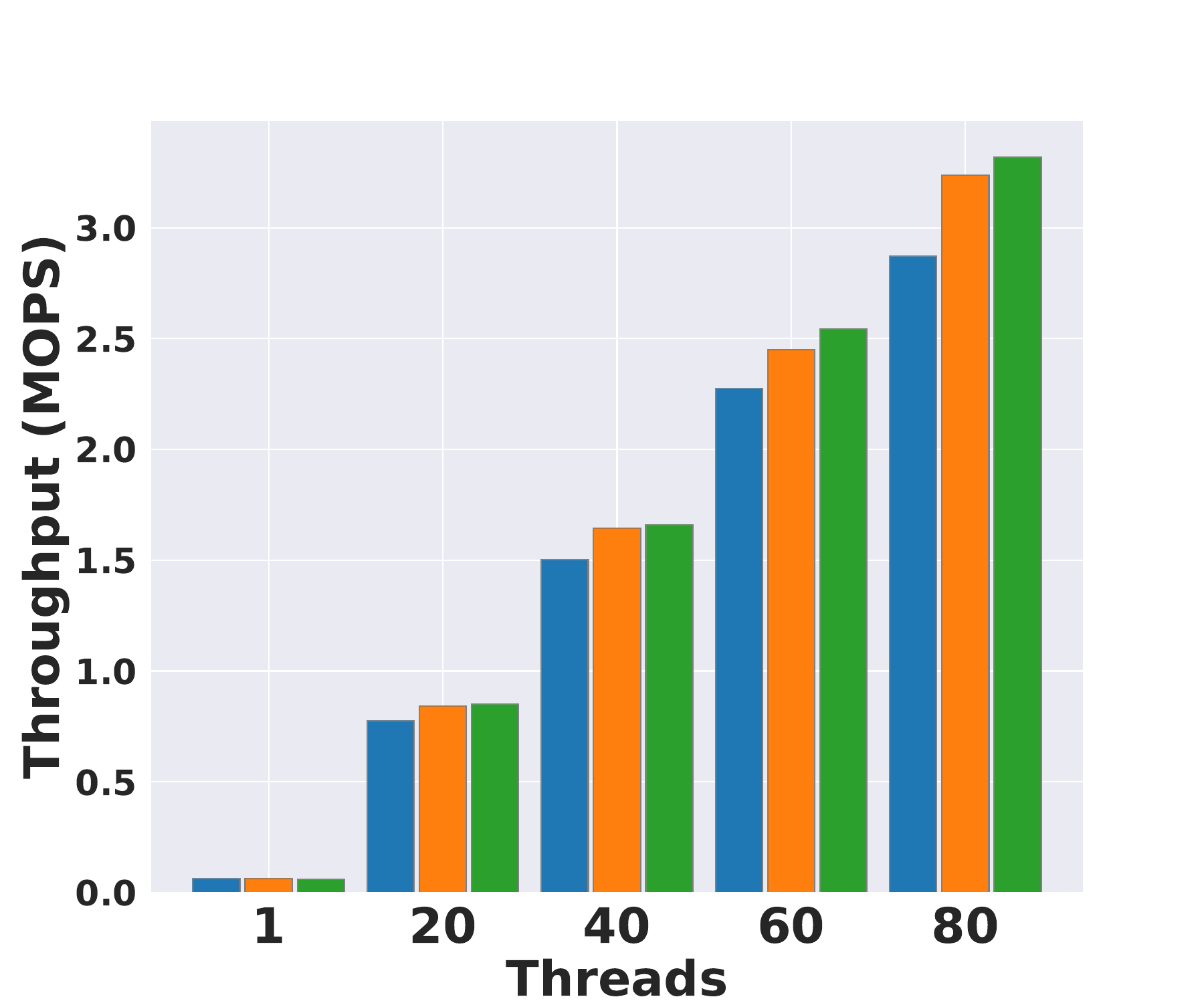} \\
   (a) & (b) & (c) \\
   \includegraphics[width=0.32\linewidth]{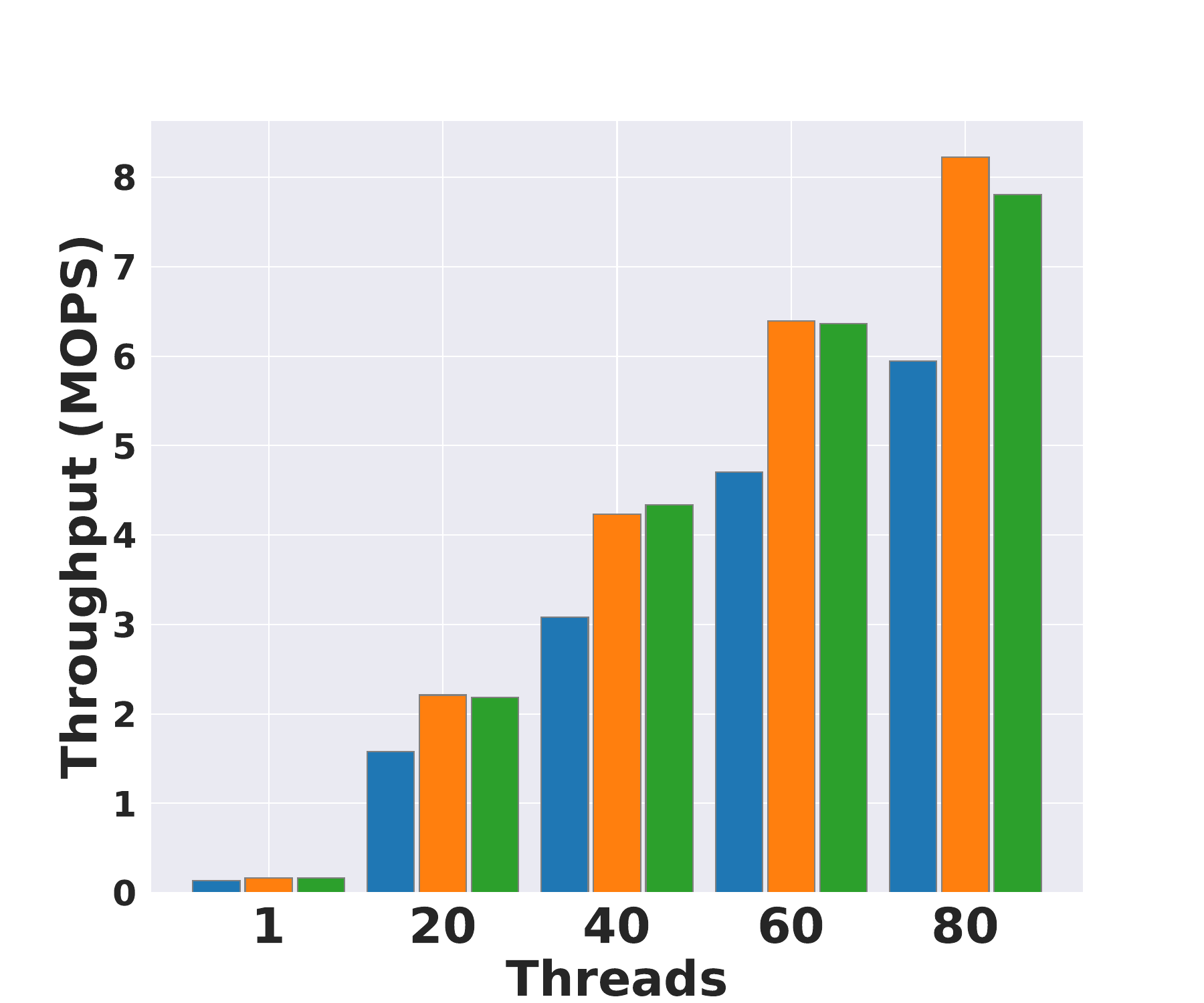} &
   \includegraphics[width=0.32\linewidth]{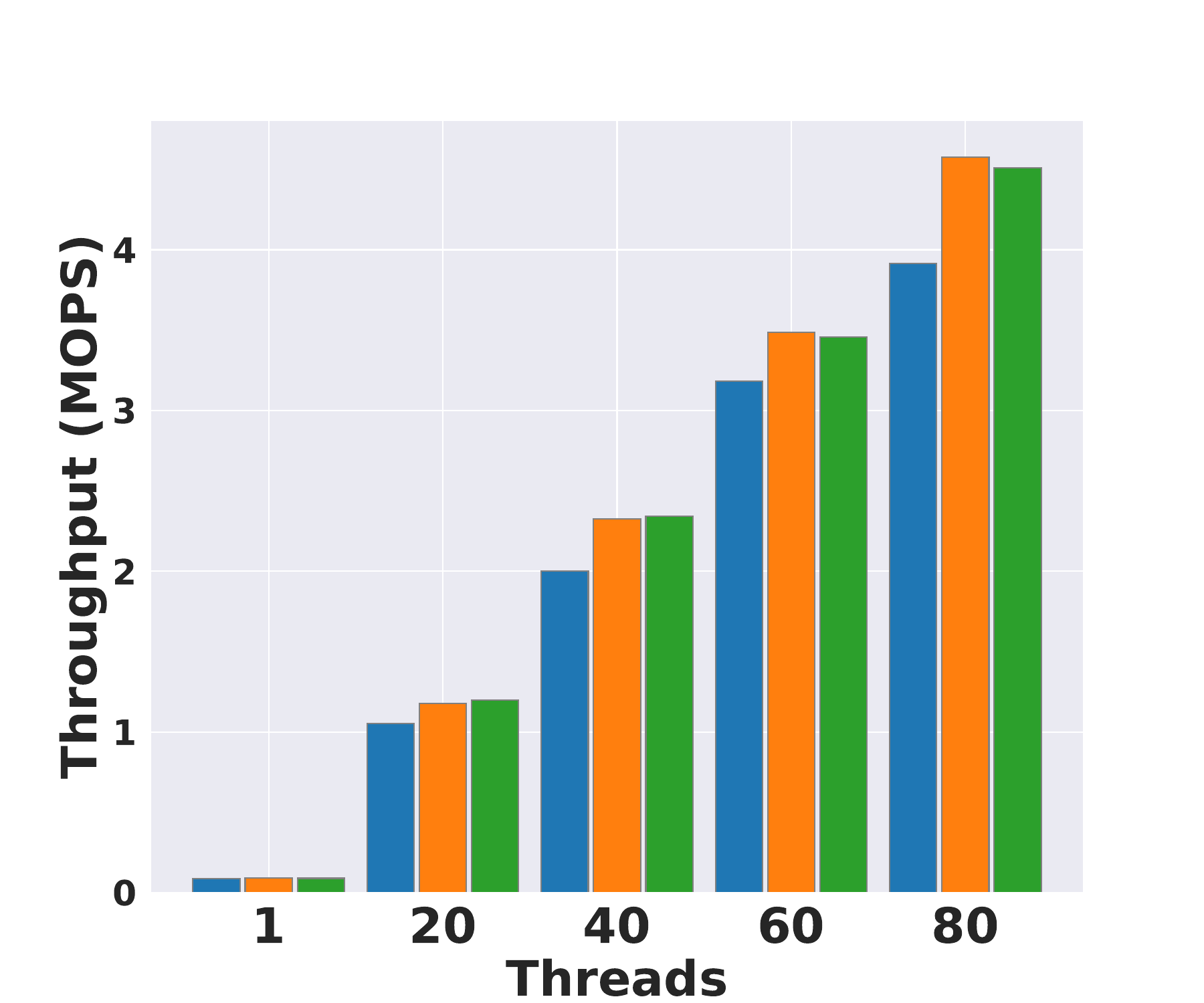} &
   \includegraphics[width=0.32\linewidth]{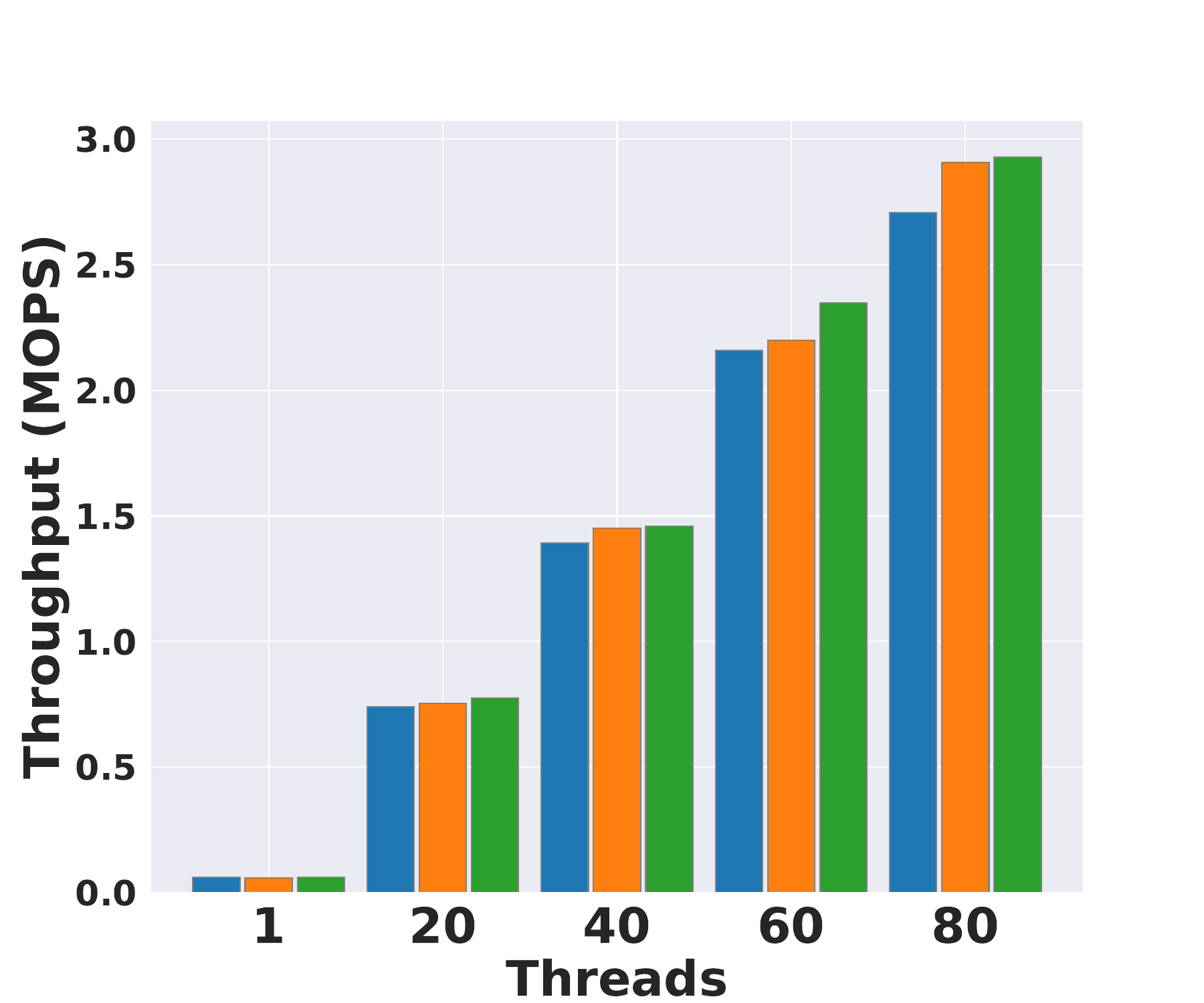} \\
   (d) & (e) & (f)
\end{tabular}
\captionof{figure}{\scriptsize The performance of \textbf{Uruv} when compared to \textbf{VCAS-BST}. The workload distributions are (a) Reads - 94\%, Updates - 5\%, Range Queries of size 1K - 1\%, (b) Reads - 90\%, Updates - 5\%, Range Queries of size 1K - 5\%, (c) Reads - 85\%, Updates - 5\%, Range Queries of size 1K - 10\%, (d) Reads - 49\%, Updates - 50\%, Range Queries of size 1K - 1\%, (e) Reads - 45\%, Updates - 50\%, Range Queries of size 1K - 5\%, and (f) Reads - 40\%, Updates - 50\%, Range Queries of size 1K - 10\%}
\label{fig:5d-5f}

\paragraph{Benchmark.} Our benchmark takes 7 parameters - read, insert, delete, range query, range query size, prefilling size, and dataset size. Read, insert, delete, and range queries indicate the percentage of these operations. We use a uniform distribution to choose between these four operations probabilistically. We prefill each data structure with 100 million keys, uniformly at random, from a universe of 500 million keys ranging [1, 500M].

\paragraph{Performance for dictionary operations.} Results of three different workloads - Read-only(Fig. \ref{fig:uruv-anastasia-openbwtree}a), Read-Heavy(Fig. \ref{fig:uruv-anastasia-openbwtree}b), and a Balanced workload(Fig. \ref{fig:uruv-anastasia-openbwtree}c) are shown in Figure \ref{fig:uruv-anastasia-openbwtree}. Across the workloads, at 80 threads, Uruv beats LF$\_$B$^+$Tree\cite{braginsky+:SPAA:2012} by \textbf{95x}, \textbf{76x}, and \textbf{44x} as it replaces the node with a new node for every insert. Uruv beats OpenBwTree\cite{wang2018building} by \textbf{1.7x}, \textbf{1.7x}, and \textbf{1.25x}. 
The performance of LF-URUV and WF-URUV correlates since WF-URUV has a lower possibility of any thread taking a slow path. In all three cases, the gap between Uruv and the rest increases as the number of threads increases. This shows the scalability of the proposed method. As we move from 1 to 80 threads, Uruv scales \textbf{46x} to \textbf{61x} in performance, LFB$^+$Tree scales \textbf{2.4x} to \textbf{5x} and OpenBw-Tree scales \textbf{39x} to \textbf{42x}. These results establish the significantly superior performance of Uruv over its existing counterpart.

\paragraph{Performance for workloads including range search.} We compare Uruv against VCAS-BST in various workloads in Figure \ref{fig:5d-5f}. Figures \ref{fig:5d-5f}a - \ref{fig:5d-5f}c are read-heavy workloads and \ref{fig:5d-5f}d - \ref{fig:5d-5f}f are update-heavy workloads. Across each type of workload, we vary the range query percentage from 1\% to 10\%. At 80 threads, we beat VCAS-BST by \textbf{1.38x} in update-heavy workloads and \textbf{1.68x} in read-heavy workloads. 
These set of results demonstrate the efficacy of Uruv's wait-free range search.


\section{Related Work}
\label{sec:Related}
We have already discussed the salient points where Uruv differs from existing techniques of concurrent range search. In particular, in contrast to the locking method of bundled references \cite{nelson2021bundled} and the lock-free method of constant time snapshots \cite{wei+:PPOPP:2021}, Uruv guarantees wait-freedom. The architecture ensuring wait-freedom in Uruv, i.e., its \texttt{stateArray}, has to accommodate its multi-versioning. The existing methods did not have to consider this.

Anastasia et al. \cite{braginsky+:SPAA:2012} developed the first lock-free B$^+$Tree. In their design, every node implements a linked-list augmented with an array. This ensures that each node in the linked-list is allocated contiguously. It slows down updates at the leaf and traversal down their tree. Uruv’s design is inspired by their work, but, does away with the arrays in the nodes. As the experiments showed, it clearly benefits. Most importantly, we also support linearizable wait-free range search, which is not available in \cite{braginsky+:SPAA:2012}. OpenBw-Tree \cite{wang2018building} is an optimized lock-free B$^+$tree that was designed to achieve high performance under realistic workloads. However, again, it does not support range search.

We acknowledge that other recently proposed tree data structures could be faster than Uruv, for example, C-IST \cite{brown+:PPOPP2020} and LF-ABTree \cite{TrevorBrown:PhDThesis}. However, LF-ABTree is a relaxed tree where the height and the size of the nodes are relaxed whereas C-IST \cite{brown+:PPOPP2020} uses interpolation search on internal nodes to achieve high performance. That is definitely an attractive dimension towards which we plan to adapt the design of Uruv. Furthermore, they are not wait-free. Our focus was on designing a B$^+$Tree that supports wait-free updates and range search operations.

In regards to wait-free data structures, most of the attempts so far has been for Set or dictionary abstract data types wherein only insertion, deletion, and membership queries are considered. For example, Natarajan et al. \cite{Natarajan2013ConcurrentWR} presented wait-free red-black trees.  Applying techniques similar to fast-path-slow-path, which we used, Petrank and Timmet \cite{petrank2017practical} proposed converting lock-free data structures to wait-free ones. They used this strategy to propose wait-free implementations of inked-list, skip-list and binary search trees. There have been prior work on wait-free queues and stacks \cite{fatourou2011highly}, \cite{attiya2018nontrivial}. However, to our knowledge, this is the first work on a wait-free implementation of an abstract data type that supports add, remove, search and range queries.

\section{Conclusion}
\label{sec:Conclusions}
We developed an efficient concurrent data structure Uruv that supports wait-free addition, deletion, membership search and range search operations. Theoretically, Uruv offers a finite upper bound on the step complexity of each operation, the first in this setting. On the practical side, Uruv significantly outperforms the existing lock-free B$^+$Tree variants and a recently proposed linearizable lock-free range search algorithm.
\bibliography{citation}

\begin{thebibliography}{10}

\bibitem{afek1999long}
Yehuda Afek, Hagit Attiya, Arie Fouren, Gideon Stupp, and Dan Touitou.
\newblock Long-lived renaming made adaptive.
\newblock In {\em Proceedings of the eighteenth annual ACM symposium on
  Principles of distributed computing}, pages 91--103, 1999.

\bibitem{arbel2018harnessing}
Maya Arbel-Raviv and Trevor Brown.
\newblock Harnessing epoch-based reclamation for efficient range queries.
\newblock {\em ACM SIGPLAN Notices}, 53(1):14--27, 2018.

\bibitem{attiya2018nontrivial}
Hagit Attiya, Armando Casta{\~n}eda, and Danny Hendler.
\newblock Nontrivial and universal helping for wait-free queues and stacks.
\newblock {\em Journal of Parallel and Distributed Computing}, 121:1--14, 2018.

\bibitem{basin2017kiwi}
Dmitry Basin, Edward Bortnikov, Anastasia Braginsky, Guy Golan-Gueta, Eshcar
  Hillel, Idit Keidar, and Moshe Sulamy.
\newblock Kiwi: A key-value map for scalable real-time analytics.
\newblock In {\em Proceedings of the 22Nd ACM SIGPLAN Symposium on Principles
  and Practice of Parallel Programming}, pages 357--369, 2017.

\bibitem{braginsky+:SPAA:2012}
Anastasia Braginsky and Erez Petrank.
\newblock A lock-free b+ tree.
\newblock In {\em Proceedings of the twenty-fourth annual ACM symposium on
  Parallelism in algorithms and architectures}, pages 58--67, 2012.

\bibitem{TrevorBrown:PhDThesis}
Trevor Brown.
\newblock Techniques for constructing efficient lock-free data structures.
\newblock {\em CoRR}, abs/1712.05406, 2017.
\newblock URL: \url{http://arxiv.org/abs/1712.05406}, \href
  {http://arxiv.org/abs/1712.05406} {\path{arXiv:1712.05406}}.

\bibitem{brown2012range}
Trevor Brown and Hillel Avni.
\newblock Range queries in non-blocking k-ary search trees.
\newblock In {\em International Conference On Principles Of Distributed
  Systems}, pages 31--45. Springer, 2012.

\bibitem{brown+:PPOPP2020}
Trevor Brown, Aleksandar Prokopec, and Dan Alistarh.
\newblock Non-blocking interpolation search trees with doubly-logarithmic
  running time.
\newblock In {\em Proceedings of the 25th ACM SIGPLAN Symposium on Principles
  and Practice of Parallel Programming}, pages 276--291, 2020.

\bibitem{chatterjee2017lock}
Bapi Chatterjee.
\newblock Lock-free linearizable 1-dimensional range queries.
\newblock In {\em Proceedings of the 18th International Conference on
  Distributed Computing and Networking}, pages 1--10, 2017.

\bibitem{comer+:ACM1979}
Douglas Comer.
\newblock Ubiquitous b-tree.
\newblock {\em ACM Computing Surveys (CSUR)}, 11(2):121--137, 1979.

\bibitem{fatourou2011highly}
Panagiota Fatourou and Nikolaos~D Kallimanis.
\newblock A highly-efficient wait-free universal construction.
\newblock In {\em Proceedings of the twenty-third annual ACM symposium on
  Parallelism in algorithms and architectures}, pages 325--334, 2011.

\bibitem{flurry}
{flurry}.
\newblock {Flurry Analytics}.
\newblock \url{https://www.flurry.com/}, 2022.
\newblock Online; accessed May 2022.

\bibitem{harris2001pragmatic}
Timothy~L Harris.
\newblock A pragmatic implementation of non-blocking linked-lists.
\newblock In {\em Distributed Computing: 15th International Conference, DISC
  2001 Lisbon, Portugal, October 3--5, 2001 Proceedings 15}, pages 300--314.
  Springer, 2001.

\bibitem{Herlihy+:OnNatProg:opodis:2011}
Maurice Herlihy and Nir Shavit.
\newblock {On the Nature of Progress}.
\newblock In {\em {OPODIS}}, pages 313--328, 2011.

\bibitem{herlihy1990linearizability}
Maurice~P Herlihy and Jeannette~M Wing.
\newblock Linearizability: A correctness condition for concurrent objects.
\newblock {\em ACM Transactions on Programming Languages and Systems (TOPLAS)},
  12(3):463--492, 1990.

\bibitem{kogan2012methodology}
Alex Kogan and Erez Petrank.
\newblock A methodology for creating fast wait-free data structures.
\newblock {\em ACM SIGPLAN Notices}, 47(8):141--150, 2012.

\bibitem{kowalski2020high}
Thomas Kowalski, Fotios Kounelis, and Holger Pirk.
\newblock High-performance tree indices: Locality matters more than one would
  think.
\newblock In {\em 11th International Workshop on Accelerating Analytics and
  Data Management Systems}, 2020.

\bibitem{Natarajan2013ConcurrentWR}
Aravind Natarajan, Lee Savoie, and Neeraj Mittal.
\newblock Concurrent wait-free red black trees.
\newblock In {\em Safety-critical Systems Symposium}, 2013.

\bibitem{nelson2021bundled}
Jacob Nelson, Ahmed Hassan, and Roberto Palmieri.
\newblock Bundled references: an abstraction for highly-concurrent linearizable
  range queries.
\newblock In {\em Proceedings of the 26th ACM SIGPLAN Symposium on Principles
  and Practice of Parallel Programming}, pages 448--450, 2021.

\bibitem{petrank2017practical}
Erez Petrank and Shahar Timnat.
\newblock A practical wait-free simulation for lock-free data structures.
\newblock 2017.

\bibitem{tian2015latency}
Xinhui Tian, Rui Han, Lei Wang, Gang Lu, and Jianfeng Zhan.
\newblock Latency critical big data computing in finance.
\newblock {\em The Journal of Finance and Data Science}, 1(1):33--41, 2015.

\bibitem{Timnat+:WFLis:opodis:2012}
Shahar Timnat, Anastasia Braginsky, Alex Kogan, and Erez Petrank.
\newblock {Wait-Free Linked-Lists}.
\newblock In {\em {OPODIS}}, pages 330--344, 2012.

\bibitem{wang2018building}
Ziqi Wang, Andrew Pavlo, Hyeontaek Lim, Viktor Leis, Huanchen Zhang, Michael
  Kaminsky, and David~G Andersen.
\newblock Building a bw-tree takes more than just buzz words.
\newblock In {\em Proceedings of the 2018 International Conference on
  Management of Data}, pages 473--488, 2018.

\bibitem{wei+:PPOPP:2021}
Yuanhao Wei, Naama Ben-David, Guy~E Blelloch, Panagiota Fatourou, Eric Ruppert,
  and Yihan Sun.
\newblock Constant-time snapshots with applications to concurrent data
  structures.
\newblock In {\em Proceedings of the 26th ACM SIGPLAN Symposium on Principles
  and Practice of Parallel Programming}, pages 31--46, 2021.

\bibitem{zhang2015memory}
Hao Zhang, Gang Chen, Beng~Chin Ooi, Kian-Lee Tan, and Meihui Zhang.
\newblock In-memory big data management and processing: A survey.
\newblock {\em IEEE Transactions on Knowledge and Data Engineering},
  27(7):1920--1948, 2015.

\end{thebibliography}
\newpage
\appendix
\label{appendix}
\section{Additional Algorithmic Reference For Uruv}
\label{sec:additionalalgo}
This section is intended for an algorithmic reference. A lot of this section has already been explained in the main paper.

\begin{figure}[H]
\algrenewcommand\alglinenumber[1]{\scriptsize #1:} 

\newcommand{\algrule}[1][.5pt]{\par\vskip.5\baselineskip\hrule height #1\par\vskip.5\baselineskip} 
\begin{multicols}{2}
\begin{algorithmic}[1]
\renewcommand{\algorithmicprocedure}{}
\label{insert}
\algrestore{insert}
\scriptsize
\Statex
\Procedure{\textbf{Delete}}{key}
\State{\texttt{retry}:}
\State{$\textup{Node* } curr := root$}
\If{$curr =$ \textup{\texttt{nullptr}}}
    \State{\textbf{return} \textup{"Key Not Present"}} 
\EndIf
\State{$curr :=$ \textup{\texttt{balanceRoot(curr)}}}
\label{alg:del:balance_root}
\If{{\textup{!}}$\,curr$}{\textup{ \textbf{goto} \texttt{retry}}}
\EndIf
\State{$\textup{Node* } prev, child :=$ \texttt{nullptr}}
\State{int $pidx, cidx$}
\While{\textup{!}$curr \rightarrow isLeaf$}
    \label{alg:del:isleaf}
    \If{$curr \rightarrow helpIdx \neq -1$}
        \label{alg2:line48}
        \label{alg:del:help}
        \State{$\textup{Node*} \, res:=$ \texttt{help}($prev, pidx, curr$)}
        \If{$res$} {$curr := res$}
        \Else { \textup{\textbf{goto} \texttt{retry}}}
        \EndIf
    \EndIf
    \State{{$cidx$ \textup{is set to the index of appropriate child based on} $key$ using Binary Search}}
    \label{alg:del:curridx}
    \State{$child := curr \rightarrow ptr[cidx]$}
    \If{$child\rightarrow isLeaf  \,\&\&\ child\rightarrow frozen$}
        \State{$\,curr \rightarrow$ \texttt{freezeInternal()}}
        \If{\textup{!}$\,curr \rightarrow$\textup{\texttt{setHelpIdx$(cidx)$}}} 
        \State{\textup{\textbf{goto} \texttt{retry}}}
        
        \EndIf
        \State{$\textup{Node*} newNode:=child\rightarrow$ \texttt{balanceLeaf}($prev, pidx, curr, cidx)$}
        \If{$newNode$} \State{\textbf{then} $curr := newNode$}
        \Else{ \textup{\textbf{goto}        \texttt{retry}}}
        \EndIf
       
    \ElsIf{\textup{!}$\,child \rightarrow isLeaf \&\&\ child \rightarrow count \leq MIN$}
    {
        \State{$\,curr \rightarrow$ \texttt{freezeInternal()}}
        \If{\textup{!}$\,curr\rightarrow$\textup{\texttt{setHelpIdx$(cidx)$}}} {
         \State{\textup{\textbf{goto} \texttt{retry}}}
        }
        \EndIf
        \State{$\textup{Node* } newNode := child \rightarrow$ \texttt{mergeInternal}($prev, pidx, curr, cidx)$}
        \If{$newNode$} \State{\textbf{then} $curr := newNode$}
        \Else{ \textup{\textbf{goto}        \texttt{retry}}}
        \EndIf
    }
    \EndIf
     
    \State{$prev := curr$}
    \State{$curr := child$}
    \State{$pidx := cidx$}

\EndWhile

\State{$res:=$ $curr \rightarrow$ \textup{\texttt{deleteLeaf}}($key$)}
\If{$res = $ \textup{Failed}}
    \State{\textup{\textbf{goto}        \texttt{retry}}}
\Else{ \textbf{return} \textup{res}}
\EndIf

\EndProcedure
\algstore{delete}
\end{algorithmic}
\end{multicols}
\caption{Pseudocode of \remove operation}
\label{alg:insert}
\end{figure}

\ignore{
\begin{figure}[H]
\algrenewcommand\alglinenumber[1]{\scriptsize #1:}
\begin{multicols}{2}
\begin{algorithmic}[1]
\renewcommand{\algorithmicprocedure}\scriptsize
\algrestore{rangequery}
\Procedure{\textbf{insertLeaf}}{key,value}

\If{$frozen$}{ \textbf{return} \texttt{"Failed"}}
\EndIf
\label{alg5:line97}
\If{$count \geq MAX$}
    \State{\texttt{freezeLeaf()}}
    \label{alg5:line99}
    \State{\textbf{return}
    \texttt{"Failed"}}
    \label{alg5:line100}
\Else
    \State{res := \texttt{\textup{llInsert}}($key,\ value$)}
    \If{res = "New Key Inserted"}
    \label{alg5:line101}
    \State{$count$++}
    \EndIf
    \State{\textbf{return} res}
    
\EndIf
\EndProcedure
\algstore{insertleaf}
\end{algorithmic}
\hrule
\begin{algorithmic}[1]
\renewcommand{\algorithmicprocedure}\scriptsize
\algrestore{insertleaf}
\Procedure{\textbf{deleteLeaf}}{key}
\State{$\textup{llNode*} node :=$ \texttt{find}($key$)}
\label{alg9:line130}
\If{$node \rightarrow key = key$}
    \State{res := \texttt{insert}($key$, tombstone)}
    \label{alg9:line132}
    \State{\textbf{return} \texttt{res}}
\EndIf
\State{\textbf{return} \texttt{"Key Not Present"}}
\EndProcedure
\algstore{deleteleaf}
\end{algorithmic}
\hrule
\begin{algorithmic}[1]
\renewcommand{\algorithmicprocedure}\scriptsize
\algrestore{deleteleaf}
\Procedure{\textbf{balanceRoot}}{currRoot}
\State{$\textup{Node* } newNode$}
\If{$currRoot \rightarrow isLeaf\ \&\&\ currRoot \rightarrow frozen$}
     \If{$currRoot \rightarrow count \geq MAX$} {$newNode :=$ \texttt{splitRootLeaf}()}
    \Else { $newNode :=$ \texttt{createNewLeaf}()}
    \EndIf
\ElsIf {\textup{!}$currRoot \rightarrow isLeaf\ \&\&\ currRoot \rightarrow count \geq MAX$}{ $newNode :=$ \texttt{splitRootInternal}()}
\EndIf
\If{\textup{!}$\,root$.\texttt{CAS}($\textit{\textit{\textbf{this}}},\  newNode$)} {\textup{\textbf{return}} \texttt{nullptr}}
\Else{ \textbf{\textup{return}} $newNode$}
\EndIf
\EndProcedure
\algstore{balanceroot}
\end{algorithmic}
\hrule

\begin{algorithmic}[1]
\renewcommand{\algorithmicprocedure}\scriptsize
\algrestore{balanceroot}
\Procedure{\textbf{balanceLeaf}}{prevNode,\ prevIdx,\ currNode,\ currIdx}

\If{ $currNode \rightarrow helpIdx \neq$ -1} 
\State{\textbf{return} \texttt{nullptr}}
\ElsIf{\textup{!}$\,currNode \rightarrow helpIdx$.\texttt{CAS}(-1, $currIdx$)}
\State{\textbf{return} \texttt{nullptr}}
\EndIf
\State{$\textup{Node* } newNode$}
\If{$count \geq MAX$} 
    \State{$newNode :=$ \texttt{splitLeaf}($prevNode,$ $\ prevIdx,$ $\ currNode,\ currIdx$)}
    \label{alg8:line122}
\ElsIf{$count < MIN$}
    \State{$newNode :=$ \texttt{merge}($prevNode,$ $\ prevIdx,\ currNode,\ currIdx$)}
    \label{alg8:line124}
\Else
    \State{$newNode :=$ \texttt{createNewLeaf}($prevNode,$ $\ prevIdx,$ $\ currNode,\ currIdx$)}
    \label{alg8:line126}
\EndIf
\State{\textbf{return} $newNode$}

\EndProcedure
\algstore{balanceleaf}
\end{algorithmic}

\hrule
\begin{algorithmic}[1]
\renewcommand{\algorithmicprocedure}\scriptsize
\algrestore{balanceleaf}
\Procedure{\textbf{help}}{prevNode,\ prevIdx,\ currNode}
\State{$\textup{Node* } currChild := currNode \rightarrow ptr[currNode \rightarrow helpIdx]$}
\If{$currChild \rightarrow isLeaf$} \State{\textbf{return} \texttt{$currChild \rightarrow$ \texttt{balanceLeaf}($prevNode,\ prevIdx,\ currNode,\ currIdx)$}}
\ElsIf{$currChild \rightarrow count \geq MAX$} 
\State{\textbf{return} \texttt{$currChild \rightarrow$ \texttt{splitInternal}($prevNode,\ prevIdx,\ currNode,\ curr_idx)$}}
\Else 
\State{textbf{return} \texttt{$currChild \rightarrow$ \texttt{merge}($prevNode,\ prevIdx,\ currNode,\ currIdx)$}}
\EndIf
\EndProcedure
\algstore{help}
\end{algorithmic}

\hrule
\begin{algorithmic}[1]
\renewcommand{\algorithmicprocedure}\scriptsize
\algrestore{help}
\Procedure{\textbf{freezeInternal()}}{}
\For{$i := 0$ \textbf{to} $MAX$}
    \State{$ptr[i]$.\texttt{CAS}($ptr[i],\  $ \texttt{markedRef}($ptr[i]))$}
    \label{alg12:line160}
\EndFor
\State{\textit{frozen} := \texttt{True}}
\EndProcedure
\algstore{freezeInternal}

\end{algorithmic}
\hrule
\begin{algorithmic}[1]
\renewcommand{\algorithmicprocedure}\scriptsize
\algrestore{freezeInternal}
\Procedure{\textbf{freezeLeaf()}}{}
\State{llNode* $currNode := ver\_head$}
\While{$currNode \rightarrow next \neq$ \textup{\texttt{nullptr}}}
    \State{$currNode \rightarrow next$.\texttt{CAS}($currNode \rightarrow next,\ $ \texttt{markedRef}($currNode \rightarrow next$))}
    \label{alg13:line163}
    \State{$currNode \rightarrow vhead$.\texttt{CAS}($currNode \rightarrow vhead,\ $ \texttt{markedRef}($currNode \rightarrow vhead$))}
    \label{alg13:line164}
    \State{$currNode := currNode \rightarrow next$}
\EndWhile
\State{$\textit{\textbf{this}} \rightarrow count := list$.\texttt{size()}}
\State{\textit{frozen} := \texttt{True}}
\EndProcedure
\algstore{freezeleaf}
\end{algorithmic}
\end{multicols}
\caption{Pseudocode of insertLeaf, deleteLeaf, balanceRoot, balanceLeaf, help, freezeInternal and freezeLeaf}
\label{fig:freeze}
\end{figure}
\begin{figure}
\begin{multicols}{2}
\begin{algorithmic}[1]
\renewcommand{\algorithmicprocedure}{}
\scriptsize
\algrestore{borrowhelper}
\algrenewcommand\alglinenumber[1]{\scriptsize #1:}
\Procedure{\textbf{wfInsertLeaf}}{key,\ value,\ tid,\ ph}
\If{$frozen$}
\State{\textbf{return} \texttt{"Failed"}}
\EndIf
\If{$count \geq MAX$}
    \State{\texttt{freezeLeaf()}}
    \State{\textbf{return} \texttt{"Failed"}}
\Else 
\State{res := \texttt{\textup{wfLLInsert}}($key,\ value,\ tid,\ ph$)}
\If{res = \texttt{"New Key Inserted"}}
    \State{$count$++}
\EndIf
\textbf{return} \texttt{res}
\EndIf
\EndProcedure
\algstore{wfllinsert}
\end{algorithmic}
\hrule
\begin{algorithmic}[1]
\renewcommand{\algorithmicprocedure}{}
\scriptsize
\algrestore{wfllinsert}
\algrenewcommand\alglinenumber[1]{\scriptsize #1:}
\Procedure{\textbf{wfDeleteLeaf}}{key,\ tid,\ phase}
\State{$currState := stateArray[tid]$}
\State{$currNode := currState \rightarrow searchNode$}
\label{alg10:line136}
\If{$currNode = nullptr$}
    \State{$currState \rightarrow finished := true $}
    \State{\textbf{return} \texttt{"Operation Finished"}}
\EndIf
\If{$currNode = dummyNode$}
    \State{$node := list$.\texttt{find}$(key)$}
    \label{alg11:find}
    \State{$currState \rightarrow searchNode.$\texttt{CAS} $(dummyNode, Node)$}
    \label{alg10:line142}
\EndIf
\State{$\textup{llNode*} \ node := currState \rightarrow searchNode$}
\If{$node == nullptr$}
    \State{$currState \rightarrow finished := true $}
    \State{\textbf{return} \texttt{"Operation Finished"}}
\EndIf
\If{$node \rightarrow key = key$}
    \State{\texttt{wfLLInsert}($key$, -1,\ tid,\ phase)}
    \label{alg10:line151}
    \State{$currState \rightarrow finished := true$}
    \State{\textbf{return} \texttt{"Operation Finished"}}
\Else
    \State{$currState \rightarrow finished := true$}
    \State{\textbf{return} \texttt{"Operation Finished"}}
\EndIf
\EndProcedure
\algstore{wfdeleteleaf}
\end{algorithmic}
\end{multicols}
\caption{Pseudocode of wfInsertLeaf snd wfDeleteLeaf \op}
\end{figure}
\begin{algorithm}[H]
\label{balanceroot}
\caption{\texttt{balanceRoot}()}
$\textup{Node* } newNode$\;
\uIf{$isLeaf\ \&\&\ frozen$}
{
    \textbf{if} {$count \geq MAX$} \textbf{then} $newNode :=$ \texttt{splitRootLeaf}()\;
    \textbf{else} $newNode :=$ \texttt{createNewLeaf}()\;
}
\textbf{else if} {\textup{!}$\,isLeaf\ \&\&\ count \geq MAX$} \textbf{then} $newNode :=$ \texttt{splitRootInternal}()\;
\textbf{if} {\textup{!}$\,root$.\texttt{CAS}($\textit{\textit{\textbf{this}}},\  newNode$)} \textbf{then} \textup{\textbf{return}} \texttt{nullptr}\;
\textbf{else} \textbf{\textup{return}} $newNode$\;
\end{algorithm}
\begin{algorithm}[H]
\label{splitrootleaf}
\caption{\texttt{splitRootLeaf}}
$\textup{Node* } leftChild :=$ \texttt{new leafNode()}\;
$\textup{Node* } rightChild :=$ \texttt{new leafNode()}\;
$leftChild \rightarrow next := rightChild$ \;
$rightChild \rightarrow next := next$ \;
Split \textit{\textbf{this}} node into a pair of nodes pointed to by $leftChild$ and $rightChild$\;
\If{\textup{!}$\ newNext$.\texttt{\textup{CAS}}($nullptr,\ leftChild$)}{
    $leftChild := newNext$\;
    $rightChild := leftChild \rightarrow next$\;
}

$\textup{Node* } newNode :=$ \texttt{new internalNode()}\;
$newNode \rightarrow ptr[0] := leftChild$\;
$newNode \rightarrow ptr[1] := rightChild$\;
$newNode \rightarrow key[0] := rightChild's$  first key\;
\textbf{if} \textup{!}$\,root$.\texttt{CAS}($this,\ newNode$) \textbf{then} \textbf{return} \texttt{nullptr}\; 
\textbf{return} $newNode$\;
\end{algorithm}
\begin{algorithm}[H]
\label{splitrootinternal}
\caption{\texttt{splitRootInternal}}
\texttt{markInternal()}\;
$\textup{Node* } leftChild :=$ \texttt{new internalNode()}\;
$\textup{Node* } rightChild :=$ \texttt{new internalNode()}\;
Split \textbf{\textit{this}} node into a pair of nodes pointed to by $leftChild$ and $rightChild$\;
$\textup{Node* } newNode :=$ \texttt{new internalNode()}\;
$newNode \rightarrow ptr[0] := leftChild$\;
$newNode \rightarrow ptr[1] := rightChild$\;
$newNode \rightarrow ptr[1] := $ first key from the second half of the current list of keys\;
\textbf{if} \textup{!}$\,root$.\texttt{CAS}($this,\ newNode$) \textbf{then} \textbf{return} \texttt{nullptr}\;
\textbf{return} $newNode$\;
\end{algorithm}
\begin{algorithm}[H]
\label{balanceleaf}
\caption{\texttt{balanceLeaf}(prevNode,\ prevIdx,\ currNode,\ currIdx)}
$\textup{Node* } newNode$\;
\uIf{$count \geq MAX$} 
{
    $newNode :=$ \texttt{splitLeaf}($prevNode,\ prevIdx,\ currNode,\ currIdx$)\;
    \label{alg8:line122}
}
\uElseIf{$count < MIN$}
{
    $newNode :=$ \texttt{merge}($prevNode,\ prevIdx,\ currNode,\ currIdx$)\;
    \label{alg8:line124}
}
\Else
{
    $newNode :=$ \texttt{createNewLeaf}($prevNode,\ prevIdx,\ currNode,\ currIdx$)\;
    \label{alg8:line126}
}
\textbf{return} $newNode$\;
\end{algorithm}
\begin{algorithm}[H]
\label{deletefromleaf}
\caption{\texttt{deleteFromLeaf}(key)}
\textbf{if} $frozen$ \textbf{then} \textbf{return} \texttt{false}\;
$\textup{llNode*} \ node := list$.\texttt{llFind}($key$)\;
\label{alg9:line130}
\If{$node \rightarrow key = key$}
{
    \textbf{return} $list$.\texttt{llInsert}($key$, -1)\;
    \label{alg9:line132}
}
\textbf{else} \textbf{return} \texttt{true}\;
\end{algorithm}

\begin{algorithm}[H]
\label{wfdeletefromleaf}
\caption{\texttt{wfDeleteFromLeaf}(key,\ tid,\  phase)}
\textbf{if} $frozen$ \textbf{then} \textbf{return} \texttt{false}\;
$currState := stateArray[tid]$\;
$currNode := currState \rightarrow searchNode$\;
\label{alg10:line136}
\If{$currNode == nullptr$}{
    $currState \rightarrow finished := true $\;
    \textbf{return} \texttt{true}\;
}
\If{$currNode == dummyNode$}{
    $node := list$.\texttt{llFind}$(key)$\;
    \label{alg11:find}
    $currState \rightarrow searchNode.$\texttt{CAS}$(dummyNode, Node)$\;
    \label{alg10:line142}
}
$\textup{llNode*} \ node := currState \rightarrow searchNode$\;
\If{$node == nullptr$}{
    $currState \rightarrow finished := true $\;
    \textbf{return} \texttt{true}\;
}
\uIf{$node \rightarrow key == key$}
{
    \texttt{bool} $res := list$.\texttt{wfLLInsert}($key$, -1,\ tid,\ phase)\;
    \label{alg10:line151}
    \If{res}{$currState \rightarrow finished := true$\;}
    \textbf{return} \texttt{res}\;
}
\Else{
    $currState \rightarrow finished := true$\;
    \textbf{return} \texttt{true}\;
}
\end{algorithm}

\begin{algorithm}[H]
\label{dohelping}
\caption{\texttt{help}(prevNode,\ prevIdx,\ currNode)}
$\textup{Node* } currChild := currNode \rightarrow ptr[currNode \rightarrow helpIdx]$\;
\textbf{if} $currChild \rightarrow isLeaf$ \textbf{then} \textbf{return} \texttt{$currChild \rightarrow$ \texttt{balanceLeaf}($prevNode,\ prevIdx,\ currNode,\ currIdx)$}\;
\textbf{else if} $currChild \rightarrow count \geq MAX$ \textbf{then} \textbf{return} \texttt{$currChild \rightarrow$ \texttt{splitInternal}($prevNode,\ prevIdx,\ currNode,\ curr_idx)$}\;
\textbf{else} \textbf{return} \texttt{$currChild \rightarrow$ \texttt{merge}($prevNode,\ prevIdx,\ currNode,\ currIdx)$}\;
\end{algorithm}
\begin{algorithm}[H]
\label{markinternal}
\caption{\texttt{freezeInternal()}}
\For{$i := 0$ \textbf{to} $MAX$}
{
    
    $ptr[i]$.\texttt{CAS}($ptr[i],\  $ \texttt{markedRef}($ptr[i]))$\;
    \label{alg12:line160}
}
$frozen := true$\;
\end{algorithm}
\begin{algorithm}[H]
\label{markleaf}
\caption{\texttt{freezeLeaf()}}
llNode* $currNode := list.head$\;
\While{$currNode \rightarrow next \neq$ \textup{\texttt{nullptr}}}
{
    $currNode \rightarrow next$.\texttt{CAS}($currNode \rightarrow next,\ $ \texttt{markedRef}($currNode \rightarrow next$))\;
    \label{alg13:line163}
    $currNode \rightarrow vhead$.\texttt{CAS}($currNode \rightarrow vhead,\ $ \texttt{markedRef}($currNode \rightarrow vhead$))\;
    \label{alg13:line164}
    $currNode := currNode \rightarrow next$\;
}
$frozen := true$\;
$count := list$.\texttt{size()}\;
\end{algorithm}
}
\vspace{-0.2cm}\paragraph*{Freezing.} Before a node is split or merged, it is frozen along with its parent. Once the node is frozen, it becomes immutable. In an internal node, every child pointer will be marked by setting a designated bit one after the other at line \ref{alg12:line160}. Once all the child pointers are marked, the node is frozen. Unlike internal nodes, leaf nodes are frozen by marking all the linked-list entries one after the other. Nodes in the linked list are marked by marking the two pointers inside them. First, they are marked by setting the designated bit in the \textit{next} pointer at line \ref{alg13:line163}. After this, no more nodes can be added or deleted from the linked list.  Once this is done, the \textit{vhead}s are also marked in the linked-list nodes ensuring that no more updates can happen at any linked-list node at line \ref{alg13:line164}. The linked list becomes immutable once they are all marked, thereby freezing the leaf node. For more algorithmic details, please refer to \texttt{freezeInternal} and \texttt{freezeLeaf} in Fig \ref{fig:freeze}.

\vspace{-0.2cm}\paragraph*{Balancing a Leaf.} In this method leaf node is balanced by splitting or merging based on the number of elements in the node. The parent node, $currNode$, is already frozen, which prevents any thread from updating it. If the number of elements in the leaf is less than the minimum threshold, we either borrow a key from its sibling or merge it at line \ref{alg8:line124}. If the node has more elements than the maximum threshold, we split it at line \ref{alg8:line122}. The count of the leaf node can be inconsistent, so we might not need to split or merge the leaf. Since the node is frozen and cannot be used further, we replace it with a new copy at line \ref{alg8:line126}. Please refer to Figure \ref{fig:freeze} for more algorithmic details.
\begin{figure}[H]
\algrenewcommand\alglinenumber[1]{\scriptsize #1:}
\begin{multicols}{2}
\begin{algorithmic}[1]
\renewcommand{\algorithmicprocedure}\scriptsize
\algrestore{delete}
\Procedure{\textbf{\search}}{key}
\State{Node* $currNode := root$}
\While{\textup{!}$\,currNode \rightarrow isLeaf$}
    \State{Search $currNode$ for the appropriate child based on $key$}
    \State{Set $currNode$ to said child}
\EndWhile
\State{llNode* $foundNode := currNode \rightarrow$ \texttt{llFind}($key$)}
\If{$foundNode \rightarrow key = key$}
    \State{\textbf{return} \texttt{read} $(foundNode)$}
\Else
    \State{\textbf{return} \textup{"Key Not Found"}}
\EndIf
\EndProcedure
\algstore{search}
\end{algorithmic}
\hrule
\begin{algorithmic}[1]
\renewcommand{\algorithmicprocedure}\scriptsize
\algrestore{search}
\Procedure{\textbf{setHelpIdx}}{currIdx}   
\If{$helpIdx \neq -1$} {  \textbf{return} \texttt{false}}
\ElsIf{ \texttt{!}$helpIdx$\texttt{.CAS}$(-1,\ currIdx)$} 
\label{alg17:line218}
\State{  \textbf{return} \texttt{false}}
\Else{ \textbf{return} \texttt{true}}
\EndIf
\EndProcedure
\algstore{setidx}
\end{algorithmic}
\hrule
\begin{algorithmic}[1]
\renewcommand{\algorithmicprocedure}\scriptsize
\algrestore{setidx}
\Procedure{\textbf{\rangeQuery}}{low,high}
\State{vector $\langle$ pair $\langle$ key,value$\rangle \rangle$ $res$}
\State{int $ts :=$ \texttt{addTimestamp()}}
\label{alg4:line83}
\State{Node* $currNode := root$}
\While{\textup{!}$\,currNode \rightarrow isLeaf$}
    \State{Search $currNode$ for the appropriate child based on $key$}
    \State{Set $currNode$ to said child}
\EndWhile
\State{Node* $prevNode :=$ \texttt{nullptr}}
\While{$currNode \neq$ \textup{\texttt{nullptr}}}
{
\If{$currNode \rightarrow newNext \neq$ \textup{\texttt{nullptr}} $\&\& \ currNode \rightarrow newNext \rightarrow ts \leq ts$}
        \label{alg5:line93}
        \State{$currNode := currNode \rightarrow newNext$}
        \label{alg5:line94}
        \If {$prevNode \neq$ \texttt{nullptr}} {$prevNode \rightarrow next := currNode$}
        \EndIf
    \Else
        \State{$res := currNode$.\texttt{rangeQuery}($low, high, ts, res$)}
        \State{$prevNode := currNode$} \State{$currNode := currNode \rightarrow next$}
        \If{$currNode$'s first key $> high$} \State{\textbf{return} $res$}
        \EndIf
    \EndIf
}
\EndWhile
\State{\textbf{return} $res$}
\EndProcedure
\algstore{rangequery}
\end{algorithmic}
\end{multicols}
\caption{Pseudocode of \search, setHelpIdx and \rangeQuery}
\label{sethelpidx}
\end{figure}


\vspace{-0.2cm}\paragraph*{Deleting from a leaf.} In the versioned lock-free linked list, the physical deletion of keys never happens. We will search for the key in the linked list at line \ref{alg9:line130}. If the key is not present, deletion is unnecessary; otherwise, the key is marked as deleted by updating its value to tombstone using \texttt{vCAS}. When the value of the key is updated with a tombstone, the delete is visible to other threads operating on that linked list. 

\vspace{-0.2cm}\paragraph*{Split.} Without loss of generality, let us consider the case where we are splitting a leaf node, $currLeaf$. This method creates two new leaf nodes, \textit{leftChild} and \textit{rightChild}, by splitting $currLeaf$ into two equal parts at line \ref{alg19:line233}. We then check if \textit{newNext} is set or not at line \ref{alg19:line234}. If it is, that would mean some other thread already split the leaf, and we use the children created by that thread. We then set \textit{leftChild}’s \textit{next} pointer to \textit{rightChild} at line \ref{alg19:line237}. After this, $currLeaf$’s parent will be copied into a new internal node replacing the pointer to $currLeaf$ with the pointers to the new leaf nodes at line \ref{alg19:line239}. The parent’s keys are adjusted with respect to the new children. Let us call the updated parent copy $newNode$. We now need to replace the parent with $newNode$ atomically. If the parent is the root, we try to update the root to $newNode$ via a \textbf{\texttt{CAS}} at line \ref{alg19:line241}. Otherwise, we atomically replace the pointer to the parent with a pointer to $newNode$, in the grandparent at line \ref{alg19:line242}. Please refer to Figure \ref{fig:Tree Operations}(a) for a diagrammatic example.\\

The significant difference between splitting an internal node and a leaf node is the lack of extra care needed to ensure that the leaf nodes are connected and appropriate $newNext$ pointers are set for an accurate scan of leaf nodes. The way the parent is updated is slightly different in both cases but does not warrant an explanation as these algorithmic techniques are commonplace in the B$^+$Tree literature. Please refer to Figure \ref{merge} for more algorithmic details.

\vspace{-0.2cm}\paragraph*{Merge.} Similarly, let us consider the case where we need to merge a leaf node, $currLeaf$. We first try to merge with or borrow a key from $currLeaf$’s left sibling. If it does not exist, we try with its right sibling. Considering the case when the left sibling exists, we first freeze it at line \ref{alg14:line171}. If the total size of the left sibling and $currLeaf$ is less than the maximum threshold, we merge them. If the total size is too big, we borrow a key from the left sibling.

To merge them, we combine the keys and values from both siblings into a new leaf node, $newChild$ at line \ref{alg15:line185}. If the $newNext$ pointer is already set at line \ref{alg15:line186}, we set $newChild$ to that instead. Then, we copy the keys and pointers in their parent into a new node, replacing both sibling pointers with a pointer to $newChild$. $newNode$ represents this updated parent copy at line \ref{alg15:line189}. If we decide to borrow from the left sibling instead, we remove the sibling’s last key and make it $currLeaf$’s first key. Since both siblings are frozen, they need to be copied. Then, their parent is copied, and the pointers to the siblings are replaced with the pointers to their copies. The parent’s keys are adjusted with respect to the updated children.
\begin{figure}[H]

\algrenewcommand\alglinenumber[1]{\scriptsize #1:}
\begin{multicols}{2}
\begin{algorithmic}[1]
\renewcommand{\algorithmicprocedure}{}
\label{merge}
\scriptsize
\algrestore{freezeleaf}
\Procedure{\textbf{merge}}{prev,\ pidx,\ curr,\ cidx}
\State{int $lsIdx := cidx$ - 1}
\State{int $rsIdx := cidx$ + 1}
\If{$lsIdx \geq 0$}
    \State{$\textup{Node* } lChild := curr \rightarrow ptr[lsIdx]$}
    \If{$lChild \rightarrow$ isLeaf }
    \State{$lChild \rightarrow$ freezeLeaf}
    \label{alg14:line171}
    \Else
    \State{$lChild \rightarrow$ freezeInternal}
    \EndIf
    
    \State{\textbf{return} \texttt{mergeHelper}($\textit{\textbf{this}},$ $\ cidx,\ leftChild,$ $ lsIdx,\ $ $ prev,\ pidx,\ curr)$}
\ElsIf{$rsIdx \leq curr \rightarrow count$}
    \State{$\textup{Node* } rChild := curr \rightarrow ptr[rsIdx]$}
    \If{$lChild \rightarrow$ isLeaf}
    \State{$rChild \rightarrow$ freezeLeaf}
    \label{alg14:line175}
    \Else
    \State{$rChild \rightarrow$ freezeInternal}
    \EndIf
    \State{\textbf{return} \texttt{mergeHelper}($rChild, $ $\ rsIdx,\ \textit{\textbf{this}}, $ $\ cidx,\ prev,$ $ \ pidx,\ curr)$}

\EndIf
\EndProcedure
\algstore{merge}
\end{algorithmic}
\hrule 
\begin{algorithmic}[1]
\renewcommand{\algorithmicprocedure}{}
\scriptsize
\algrestore{merge}
\Procedure{\textbf{splitInternal}}{prevNode,\ prevIdx,\ currNode,\ currIdx}
\State{\texttt{markInternal()}}
\State{$\textup{Node* } lChild :=$ \texttt{new internalNode()}}
\State{\textup{Node* } rChild := \texttt{new internalNode()}}
\State{Split \textbf{\textit{this}} node into a pair of nodes pointed to by $lChild$ and $rChild$}
\State{$\textup{Node* } newNode :=$ \texttt{new internalNode()}}
\State{Copy elements from $currNode$ to $newNode$ replacing $currChild$ with the two pointers $lChild$ and $rChild$}
\If{$prevNode =$ \textup{\texttt{nullptr}}}
    \If{\textup{!}$\,root$.\texttt{CAS}($currNode,\ newNode$)} \State{\textbf{return} \texttt{nullptr}}
    \EndIf 
\ElsIf {!$\,prevNode \rightarrow ptr[prevIdx]$.\texttt{CAS} ($currNode, newNode$)} 
\State{\textbf{return} \texttt{nullptr}}
\EndIf
\State{\textbf{return} $newNode$}
\EndProcedure
\algstore{splitinternal}
\end{algorithmic}

\hrule
\begin{algorithmic}[1]
\renewcommand{\algorithmicprocedure}{}
\scriptsize
\algrestore{splitinternal}
\Procedure{\textbf{splitLeaf}}{prevNode,\ prevIdx,\ currNode,\ currIdx}

\State{$\textup{Node* } lChild :=$ \texttt{new leafNode()}}
\State{$\textup{Node* } rChild :=$ \texttt{new leafNode()}}
\State{Split \textit{\textbf{this}} node into a pair of nodes pointed to by $lChild$ and $rChild$}
\label{alg19:line233}
\If{\textup{!}$newNext$.\texttt{\textup{CAS}}($nullptr,lChild$)}
    \label{alg19:line234}
    \State{$lChild := newNext$}
    \State{$rChild := leftChild \rightarrow next$}
\EndIf
\State{$lChild \rightarrow next := rChild$}
\label{alg19:line237}
\State{$\textup{Node* } newNode :=$ \texttt{new internalNode()}}
\State{Copy elements from $currNode$ to $newNode$ replacing $currChild$ with the two pointers $lChild$ and $rChild$}
\label{alg19:line239}
\If{$prevNode =$ \textup{\texttt{nullptr}}}
    \If{\textup{!}$\,root$.\texttt{CAS}($currNode,\ newNode$)}
    \State{\textbf{return} \texttt{nullptr}} \label{alg19:line241}
    \EndIf
\ElsIf{!$\,prevNode \rightarrow ptr[prevIdx]$.\texttt{CAS} ($currNode,\ newNode$)}
\State{\textbf{return} \texttt{nullptr}} 
\label{alg19:line242}
\EndIf
\State{\textbf{return} $newNode$}

\EndProcedure
\algstore{splitLeaf}
\end{algorithmic}

\end{multicols}
\caption{Pseudocode of merge, splitInternal and splitLeaf \op}
\label{merge}
\end{figure}

If their parent is the root, we try to update the root to $newNode$ via a \textbf{\texttt{CAS}} at \ref{alg15:line194}. Otherwise, we atomically replace the pointer to our parent with a pointer to $newNode$ in our grandparent at line \ref{alg15:line195}. Please refer to Figure \ref{fig:Tree Operations}(b) for a diagrammatic example. Notice that while merging leaf nodes, we need extra care to connect all new leaf nodes and appropriate $newNext$ pointers are set for correct range scans. Please refer to Figure \ref{merge} for more algorithmic details.
\vspace{-0.2cm}\paragraph*{Thread Helping.} To make our data structure lock-free, we do helping. Before performing any split or merge operation, each thread first tries to atomically update the variable, $helpIdx$, using \cas with the index of the child on which the operation will be performed at line \ref{alg17:line218}. If $helpIdx$ is not updated successfully, then another thread has already set it to a child, and this thread will be redirected to help that child’s balancing operation. So by keeping track of the index that needs to be helped, we ensure threads help each other to complete the rebalancing of the data structure.
\begin{figure}
\begin{algorithmic}[1]
\renewcommand{\algorithmicprocedure}{}
\scriptsize
\algrestore{splitLeaf}
\algrenewcommand\alglinenumber[1]{\scriptsize #1:}
\Procedure{\textbf{mergeHelper}}{rightChild,\ rightIdx,\ leftChild,\ leftIdx,\ prev,\ pidx,\ curr}
\label{mergehelper}
\State{\textup{Node*} $newNode$}
\If{$rightChild \rightarrow count + leftChild \rightarrow count < MAX$}
    \State{\textup{Node*} $newChild$}
    \If{$leftChild \rightarrow isLeaf$}
        \State{$newChild :=$ \texttt{new LeafNode()}}
        \State{$newChild \rightarrow next := rightChild \rightarrow next$}
    \Else
        \State{$newChild :=$ \texttt{new InternalNode()}}
    \EndIf
    \State{Merge $rightChild$ and $leftChild$ into $newChild$}
    \label{alg15:line185}
    \If{\textup{!}$\,leftChild \rightarrow newNext$.\texttt{\textup{CAS}}(\texttt{\textup{nullptr}}, $newChild$)}
        \label{alg15:line186}
        \State{$newChild := leftChild \rightarrow newNext$}
    \EndIf
    \State{$newNode :=$ \texttt{new InternalNode()}}
    \State{Copy every element in $curr$ into $newNode$ replacing pointers $leftChild$ and $rightChild$ with $newChild$}
    \label{alg15:line189}
\Else
\State{$newNode :=$ \texttt{borrowHelper}($rightChild,$ $\ rightIdx,$ $\ leftChild,$ $\ leftIdx,$ $\ curr$)}
\EndIf
\If{$prev =$ \texttt{\textup{nullptr}}}
    \If{$newNode \rightarrow count = 0$} 
        \State{$newNode := newChild$ (this occurs only when the left and right children are merged and newNode has only one child)}
    \EndIf
    \If{\textup{!}$\,root$.\textup{\texttt{CAS}}($curr, newNode$) }
    \State{ \textbf{return} \textup{\texttt{nullptr}}}
    \label{alg15:line194}
    \EndIf
\ElsIf{\textup{!}$\,prev \rightarrow ptr[pidx]$.\texttt{CAS}($curr, newNode$)}
\State{\textbf{return} \textup{\texttt{nullptr}}}
\label{alg15:line195}
\EndIf

\State{\textbf{return} $newNode$}
\EndProcedure
\algstore{mergeHelper}
\end{algorithmic}
\caption{Pseudocode of mergeHelper}
\end{figure}
\begin{figure}
\begin{algorithmic}[1]
\renewcommand{\algorithmicprocedure}{}
\algrenewcommand\alglinenumber[1]{\scriptsize #1:}
\scriptsize
\algrestore{mergeHelper}
\Procedure{\textbf{borrowHelper}}{rightChild, rightIdx, leftChild, leftIdx, currNode}
\State{\textup{Node*} $newLeftChild$}
\State{\textup{Node*} $newRightChild$}
\If{$leftChild \rightarrow isLeaf$}
    \State{$newLeftChild :=$ \texttt{new LeafNode()}}
    \State{$newRightChild :=$ \texttt{new LeafNode()}}
    \State{$newLeftChild \rightarrow next := newRightChild$}
\Else
    \State{$newLeftChild :=$ \texttt{new InternalNode()}}
    \State{$newRightChild :=$ \texttt{new InternalNode()}}
\EndIf
\State{Copy every element in $leftChild$ and $rightChild$ into $newLeftChild$ and $newRightChild$ respectively}
\If{$leftChild \rightarrow count < MIN$}
    \State{Move the first key in $newRightChild$ into the end of $newLeftChild$}
\Else
\State{Move the last key in $newLeftChild$ into the beginning of $newRightChild$}
\EndIf
\If{\textup{!}$\,leftChild \rightarrow newNext$.\textup{\texttt{CAS}}(\textup{\texttt{nullptr}}, $newLeftChild$)}
    \State{$newLeftChild := leftChild \rightarrow newNext$}
    \State{$newRightChild := rightChild \rightarrow next$}
\EndIf
\State{\textup{Node*} $newNode =$ \texttt{new InternalNode()}}
\State{Copy every element in $currNode$ into $newNode$ replacing pointers $leftChild$ and $rightChild$ with $newLeftChild$ and $newRightChild$ respectively}
\State{\textbf{return} $newNode$}
\EndProcedure
\algstore{borrowhelper}
\end{algorithmic}
\caption{Pseudocode of borrowHelper}
\end{figure}

Similarly, while traversing in insertion or deletion, we check if the $helpIdx$ is set for every node we go through at lines \ref{alg1:line13} and \ref{alg2:line48}, respectively. If it is, we help that node; otherwise, we proceed as usual. For more algorithmic details, please refer to Figure \ref{fig:freeze}.

\section{Maintenance of Invariants}
\label{sec:Invariants}

Uruv’s keys and values lie inside an extensive linked list indexed by a B$^+$Tree-like structure. Efficient indexing occurs as Uruv indexes sublists contained in leaf nodes wherein every leaf node is connected to the next. We, therefore, use two invariants,

\begin{enumerate}
    \item Each linked list inside a leaf node is sorted in ascending order by keys.
    \item Every leaf node’s elements should be lesser than its next leaf node’s elements if it exists.
\end{enumerate}

We now show that the two invariants described above are always maintained before and after the completion of every operation. Lookups and range queries maintain both invariants trivially since they do not modify Uruv. Showing invariants' maintenance reduces to showing their maintenance in inserts and deletes.

We first show why invariant one is maintained. Since inserts boil down to inserting into the linked list inside leaf nodes, the invariant one is trivially satisfied as we use Harris’ linked-list design\cite{harris2001pragmatic}. Since a delete operation also updates a key’s value with a unique value, the invariant one is satisfied as the key’s position in the linked list remains unchanged. If the linked list is the only structure that ends up being modified, the second invariant is trivially maintained. However, Insert/Delete can lead to a Split/Merge operation which affects the leaf nodes. We now show how the invariants are maintained in cases such as these.

In a split operation, a leaf node, $currLeaf$, is broken into two leaf nodes equally, splitting the elements of the original leaf node. Each half of $currLeaf$ is copied over to either half. This copy, in essence, is atomic since $currLeaf$ is frozen and cannot be modified. We ensure every key in the left half is lesser than in the right. The left half’s next pointer points to the right half, thereby connecting both halves. This maintains invariant two, wherein leaf nodes are sorted. We then atomically replace the pointer to the $currLeaf$ from its parent node with the two new pointers to both halves. We accordingly modify the parent node’s keys so we can correctly index the extra child. Invariant one is trivially satisfied since each leaf node’s linked list is just being copied over from $currLeaf$.

In a merge operation, a leaf node either borrows a key from its sibling node or merges with it into one new leaf node. Like above, we create new leaf nodes in both cases and replace the old ones atomically. Let us consider the case when a leaf node, $currLeaf$, has to borrow a key from its left sibling. We move the last key in the left sibling to the beginning of $currLeaf$. Using invariant two, the last key in the left sibling is lesser than the first key in $currLeaf$. Therefore, moving this key over to the beginning of $currLeaf$ maintains both invariants. If $currLeaf$ borrows a key from the right sibling, we will move the first key in the right sibling to the end of $currLeaf$. It is straightforward to see that both invariants are maintained.

We now show correctness for the last case of a leaf node, $currLeaf$, merging with its sibling. Without losing generality, let us say $currLeaf$ merges with its left sibling. We copy the keys from $currLeaf$ and its left sibling into a new leaf node, $newLeaf$. The copy is, in essence, atomic as both siblings are frozen. We ensure that the linked list in $newLeaf$ is sorted, maintaining an invariant one. $currLeaf$’s left sibling is connected to $newLeaf$, and $newLeaf$ is connected to $currLeaf$’s right sibling. This maintains invariant two since it was already true before the merge operation started, and it remains true as $newLeaf$ is just a sorted combination of both siblings’ keys.

\section{Race Conditions for Wait-Free Uruv}
\label{sec:wait-free-appendix}
The first race condition can arise when two threads try to modify the \textit{vhead}, knowing that \textit{vnode}’s timestamp is not set. Let us say thread $t_1$ has read the current \textit{vhead} at line \ref{alg25:line298}, and finds out the current timestamp of the \textit{vnode} is not set. Similarly, thread $t_2$ reads the same information as $t_1$ and is now performing the \texttt{wfVCAS} operation at line \ref{alg25:line304}. If thread $t_2$ succeeds in changing the \textit{vhead} at line \ref{alg26:line321}, $t_1$ will fail as the current \textit{vhead} has changed. So only one thread can replace the \textit{vhead} with \textit{vnode}.

Let us discuss another race condition. A thread $t_1$ finds that the key to be inserted does not exist. It creates a new linked-list node and will add it to the linked-list with its \textit{vhead} pointing to the \textit{vnode} at line \ref{alg25:line314}. Now, $t_1$ stalls, and another thread $t_2$ finds that the key exists, and it tries to update the node’s \textit{vhead} with the \textit{vnode}. So, when thread $t_2$ reads the current \textit{vhead} of the node at line \ref{alg25:line303}, after initializing its \textit{vnode} at line \ref{alg25:init}, thread $t_2$ will not proceed further with the operation. In the end, only one thread will be able to replace the \textit{vhead} with \textit{vnode}.

Let us consider a final race condition wherein a thread tries to set the shared \texttt{stateArray} \textit{vnode}’s \textit{nextv} in the \texttt{wfVCAS} method. Let some thread $t_1$ stall just before executing line \ref{alg26:line320}, where it was about to set \textit{vnode}’s \textit{nextv} to the \textit{vhead} it read at line \ref{alg25:line297}. Let the $vhead$ it read be some \texttt{Vnode} $v_1$. While $t_1$ stalls, two things happen in parallel. First, the \textit{vhead} gets updated to another \texttt{Vnode}, $v_2$. Second, a thread, $t_2$, changes \textit{vnode}’s \textit{nextv} to $v_2$ at line \ref{alg26:line320} and successfully updates the \textit{vhead} at line \ref{alg26:line321}. Now thread $t_1$ wakes up and changes \textit{nextv} to $v_1$ at \ref{alg26:line320}, resulting in $v_2$ not being part of the versioned list. To avoid this race condition, we atomically update the \textit{nextv} of the $vnode$ at line \ref{alg26:line320} and check if it has already been changed by some other thread or not. If it has, then we restart the operation.

\label{race}
\section{Versioned Linked-List}
\label{sec:vlfll}


When we insert a new key and its value into the liked list, we first find the predecessor and successor nodes of the key to be inserted at line \ref{alg21:line252}. Unlike Herlihy’s linked list, searching a node requires no help. We traverse the linked list until we reach the first node whose key is greater than or equal to the one we are inserting. If the key is already present, we have to update its value in the versioned list by adding a \texttt{Vnode} containing the new value at the versioned list’s head, $vhead$, atomically using \texttt{vCAS} at line \ref{alg21:line260}. If the node is not present in the linked list, we atomically add the new \texttt{llNode} between the previously found predecessor and successor nodes at line \ref{alg21:line264}. Once the \texttt{llNode} or \texttt{Vnode} is added to the linked list, we update the node's timestamp using the \texttt{initTs} method at line \ref{alg21:line265} and line \ref{alg24:line281}.

When we need to read the value of a key, say at time t, we iterate the key’s versioned list and return the first value whose timestamp is lesser than or equal to t. This ensures we return a valid value if the reading thread slows down. The linearization point of the read is just after the linearization point of the update that inserted that versioned node.

When deleting a key, unlike a regular lock-free linked list where we mark the next pointer, we instead mark the key’s versioned list head, $vhead$, by updating its value to some unique marker value. For more algorithmic details, refer to Figure \ref{llfind}, \ref{llinsert} and \ref{wfllinsert}.
\begin{figure}
\begin{multicols}{2}
\begin{algorithmic}[1]
\renewcommand{\algorithmicprocedure}{}
\scriptsize
\algrestore{wfdeleteleaf}
\algrenewcommand\alglinenumber[1]{\scriptsize #1:}
\Procedure{\textbf{llFind}}{key, prevNode}
\State{$prevNode := head$}
\While{\textup{\texttt{true}}}
    \State{llNode* $rightNode := prevNode \rightarrow next$}
    \label{alg20:line243}
    \If{\textup{\texttt{isMarked}}$(rightNode)$}
        \State{ \textbf{return}\texttt{nullptr}}
    \ElsIf{$rightNode \rightarrow key \geq key$}
        \State{\textbf{return} $rightNode$}
    \Else
        \State{$prevNode := rightNode$}
    \EndIf
\EndWhile
\EndProcedure
\algstore{llfind}
\end{algorithmic}
\hrule

  \begin{algorithmic}[1]
\renewcommand{\algorithmicprocedure}{}
\scriptsize
\algrestore{llfind}
\algrenewcommand\alglinenumber[1]{\scriptsize #1:}
\Procedure{\textbf{initTs}}{Vnode}
\If{$Vnode \rightarrow ts =$ -1}
    \State{$Vnode \rightarrow ts$.\texttt{CAS}(-1, $currTs)$}
    \label{alg22:line270}
    \EndIf
\EndProcedure
\algstore{inits}
\end{algorithmic}

\hrule

  \begin{algorithmic}[1]
\renewcommand{\algorithmicprocedure}{}
\scriptsize
\algrestore{inits}
\algrenewcommand\alglinenumber[1]{\scriptsize #1:}
\Procedure{\textbf{read}}{llNode}
\State{$currVhead := llNode \rightarrow vhead$}
\State{\texttt{initTs}$(currVhead)$}
\label{alg23:line272}
\State{\textbf{return} $currVhead \rightarrow value$}
\EndProcedure
\algstore{read}
\end{algorithmic}
\hrule
\begin{algorithmic}[1]
\renewcommand{\algorithmicprocedure}{}
\scriptsize
\algrestore{read}
\algrenewcommand\alglinenumber[1]{\scriptsize #1:}
\Procedure{\textbf{vCAS}}{currNode, oldValue, value}
\State{$currVhead :=$ \texttt{unmarkedRef}$(currNode \rightarrow vhead)$}
\State{\texttt{initTs}$(currVhead)$} 
\label{alg24:line275}
\If{$currVhead \rightarrow value \neq oldValue$}
\State{\textbf{return} \texttt{false}}
\EndIf
\If{$currVhead \rightarrow value = newValue$}
\State{\textbf{return} \texttt{true}}
\EndIf
\State{Vnode* $newVnode :=$ \texttt{new Vnode}$(value)$}
\State{$newVnode \rightarrow nextv := currVhead$}
\If{$currNode \rightarrow vhead$.\texttt{CAS}$(currVhead,$ $\ newVnode)$}
    \label{alg24:line280}
    \State{\texttt{initTs}$(newVnode)$} \label{alg24:line281}
    \State{\textbf{return} \texttt{true}}
    
\Else 
\State{\texttt{initTs}$(currNode \rightarrow vhead)$}
\State{\textbf{return} \texttt{false}}
\EndIf
\EndProcedure
\algstore{vcas}
\end{algorithmic}

\hrule
\begin{algorithmic}[1]
\renewcommand{\algorithmicprocedure}{}
\scriptsize
\algrestore{vcas}
\algrenewcommand\alglinenumber[1]{\scriptsize #1:}
\Procedure{\textbf{wfVCAS}}{currNode, oldValue, value, newVnode, nextVnode, currVhead}
\If{$currVhead \rightarrow value \neq oldValue$}
\State{\textbf{return} \texttt{false}}
\EndIf
\If{ $currVhead \rightarrow value = newValue$}
\State{\textbf{return} \texttt{true}}
\EndIf
\If{$\, !\, (newVnode \rightarrow nextv.$\texttt{CAS}$(nextVnode,$ $ currVhead))$}
\State{\textbf{return} \texttt{false}}
\label{alg26:line320}
\EndIf
\If{$currNode \rightarrow vhead$.\textup{\texttt{CAS}}$(currVhead,$ $\ newVnode)$}
    \label{alg26:line321}
    \State{\texttt{initTs}$(newVnode)$}
     \label{alg26:line319}
    \State{\textbf{return} \texttt{true}}
\Else
\State{\texttt{initTs}$(currNode \rightarrow vhead)$}
\State{\textbf{return} \texttt{false}}
\EndIf
\EndProcedure
\algstore{wfvcas}
\end{algorithmic}

\end{multicols}
\caption{Pseudocode of llFind, initTs, read, vCAS and wfVCAS}
\label{llfind}
\end{figure}

\begin{figure}
    \begin{algorithmic}[1]
\renewcommand{\algorithmicprocedure}{}
\scriptsize
\algrestore{wfvcas}
\algrenewcommand\alglinenumber[1]{\scriptsize #1:}
\Procedure{\textbf{llInsert}}{key, value}
\While{\textup{\texttt{true}}}
    \State{llNode* $prevNode :=$ \texttt{nullptr}}
    \State{llNode* $nextNode :=$ \texttt{find}($key,$ $\&prevNode$)}
    \label{alg21:line252}
    \If{$nextNode =$ \textup{\texttt{nullptr}}}
        \State{\textbf{return} \texttt{"Failed"}}
    \ElsIf{$nextNode \rightarrow key = key$}
        \While{\textup{\texttt{true}}}
            \State{$currVhead := nextNode \rightarrow vhead$}
            \If{\texttt{isMarked}($currVhead$)} 
            \State{\textbf{return} \texttt{"Failed"}}
            \EndIf
            \State{$currValue :=$ \texttt{read}($currVhead)$}
            \If{\texttt{vCAS}($nextNode,\ currValue,$ $\ value,\ currVhead $) }
            \State{\textbf{return} \texttt{"Key Updated"}}
            \label{alg21:line260}
            \EndIf
        \EndWhile
    \Else
        \State{llNode* $newNode :=$ \texttt{new llNode}($key,\ value$)}
        \State{$newNode \rightarrow next := nextNode$}
        \If{$prevNode \rightarrow next$.\textup{\texttt{CAS}}($nextNode,\ newNode$)}
            \label{alg21:line264}
            \State{\texttt{initTs}$(newNode \rightarrow vhead)$}
            \label{alg21:line265}
            \State{\textbf{return} \texttt{"New Key Inserted"}}
        \ElsIf{\texttt{isMarked($prevNode \rightarrow next$)}} \State{\textbf{return} \texttt{"Failed"}}
        \EndIf
    \EndIf
\EndWhile
\EndProcedure
\algstore{llinsert}
\caption{Pseudocode of Linked-List Insert}

\label{llinsert}
\end{algorithmic}
\end{figure}

\begin{figure}
    \begin{algorithmic}[1]
\renewcommand{\algorithmicprocedure}{}
\scriptsize
\algrestore{llinsert}
\algrenewcommand\alglinenumber[1]{\scriptsize #1:}
\Procedure{\textbf{wfLLInsert}}{key, value, tid, phase}
\While{\textup{\texttt{true}}}
    \State{llNode* $prevNode :=$ \texttt{nullptr}}
    \State{llNode* $nextNode :=$ \texttt{find}($key,\  \&prevNode$)}
    \If{$nextNode ==$ \textup{\texttt{nullptr}}}
        \State{\textbf{return} \texttt{"Failed"}}
    \ElsIf{$nextNode \rightarrow key = key$}
        \While{\textup{\texttt{true}}}
            \State{$currState := stateArray[tid]$}
            \label{alg25:line294}
            \If{$currState \rightarrow finished$}
            \State{\textbf{return} \texttt{"Operation Finished"}} \label{alg25:line295}
            \EndIf
            \State{$newVnode := currState \rightarrow vNode$}
            \label{alg25:line296}
            \State{$nextVnode := newVnode \rightarrow nextv$}
            \label{alg25:line297}
            \State{$currVhead := nextNode \rightarrow vhead$}
            \label{alg25:line298}
             \State{\texttt{initTs}($currVhead)$}
             \label{alg25:init}
             \State{$currValue := currVhead \rightarrow value$}
            \label{alg25:line303}
            
            \If{$phase \neq currState \rightarrow phase\  \&\& \ currState \rightarrow vnode \rightarrow ts \neq$ -1}
                \label{alg25:line300}
                \State{$currState \rightarrow finished =$ \texttt{true}}
                \State{\textbf{return} \texttt{"Operation Finished"}}
                \label{alg25:line302}
            \EndIf
            \If{\texttt{isMarked}($currVhead$)}
            \State{\textbf{return} \texttt{"Failed"}}
            \EndIf
            \If{\textup{\texttt{wfVCAS}}($nextNode,\ currValue,\ value,\ newVnode,\
            nextVnode,\ currVhead  $)}
            
            \label{alg25:line304}
            \State{\textbf{return} \texttt{"Key Updated"}}
            \EndIf
        \EndWhile
    \Else
        \State{$currState := stateArray[tid]$}
        \If{$currState \rightarrow finished$}
        \State{\textbf{return} \texttt{"Operation Finished"}} \label{alg25:line308}
        \EndIf
        \If{$phase \neq currState \rightarrow phase\  \&\& \ currState \rightarrow vnode \rightarrow ts \neq$ -1}
                \label{alg25:line309}
                \State{$currState \rightarrow finished =$ \texttt{true}}
                \State{\textbf{return} \texttt{"Operation Finished"}}
        \EndIf    
        \State{llNode* $newNode :=$ \texttt{new llNode}($key,\ value,\ currState \rightarrow vnode$)}
        \State{$newNode \rightarrow next := nextNode$}
        \If{$prevNode \rightarrow next$.\textup{\texttt{CAS}}($nextNode,\ newNode$)}
        \label{alg25:line314}
            \State{\texttt{initTs}$(newNode \rightarrow vhead)$}
            \label{alg25:line312}
            \State{$currState \rightarrow finished := true$}
            \State{\textbf{return} \texttt{"New Key Inserted"}}
        
        \ElsIf{\texttt{isMarked}($prevNode \rightarrow next$)}
        \State{\textbf{return} \texttt{"Failed"}}
        \EndIf
    \EndIf
\EndWhile
\EndProcedure
\algstore{wfllinsert}

\end{algorithmic}
\caption{Pseudocode of wait-free Linked-List Insert}
\label{wfllinsert}
\end{figure}

\section{Version Tracker}
\label{sec:version_tracker}

\begin{figure}[!thb]	
	\begin{multicols}{2}
 \scriptsize
\begin{lstlisting}
class TrackerNode{
    int ts;
    atomic<TrackerNode*> next;
    bool finished;
}
\end{lstlisting}
\begin{lstlisting}
class TrackerList{
    TrackerNode* head;
    TrackerNode* tail;
    TrackerNode* sentinelLast;
}
\end{lstlisting}
\end{multicols}
\caption{Version Tracker Data Structure}
\label{fig:version_tracker}
\end{figure}


We keep track of active range queries to make our data structure memory-efficient. Every time we merge or split, we physically remove deleted keys that are not required in the future. Deleted keys with a timestamp lower than the timestamp of the oldest active range query can be safely removed from a leaf node’s linked list. This is because every active range query will not include the key in their result, as the key was deleted before they started. To keep track of the times every active range query started, we built a data structure called Version Tracker.

\begin{figure}
\begin{algorithmic}[1]
\renewcommand{\algorithmicprocedure}{}
\scriptsize
\algrestore{wfllinsert}
\algrenewcommand\alglinenumber[1]{\scriptsize #1:}
\Procedure{\textbf{removeHead()}}{}
\label{removehead}
\State{$currHead := head$}
\While{$currHead \rightarrow finish$}
    \State{$head$.\texttt{CAS}$(currHead,\ currHead \rightarrow next)$}
    \State{$currHead := head$}
\EndWhile
\State{\textbf{return} $currHead$}
\EndProcedure
\algstore{removehead}
\end{algorithmic}
\hrule 

\begin{algorithmic}[1]
\renewcommand{\algorithmicprocedure}{}
\scriptsize
\algrestore{removehead}
\algrenewcommand\alglinenumber[1]{\scriptsize #1:}
\Procedure{\textbf{addTimestamp()}}{}
\While{\textup{\texttt{true}}}
    \State{$currTail := tail$}
    \State{TrackerNode* $\ newTail :=$ \texttt{new TrackerNode} $(currTail \rightarrow ts + 1 ,\ sentinelLast)$}
    \If{$currTail \rightarrow next$.\textup{\texttt{CAS}} $(sentinelLast,\ newTail)$}
        \State{$tail$.\texttt{CAS}$(currTail,\ newTail)$}
        \State{\textbf{return} $newTail \rightarrow ts$}
    
    \Else
        \While{\textup{\texttt{true}}}
            \State{$currTail := tail$}
            \If{$currTail \rightarrow next = sentinelLast$} \State{\textbf{return}}
            \Else
            \State{$tail$.\texttt{CAS}$(currTail,\ currTail \rightarrow next)$}
            \EndIf
        \EndWhile
    \EndIf

\EndWhile
\EndProcedure
\end{algorithmic}
\caption{Pseudocode of removeHead and addTimestamp}
\label{alg:remHead}
\end{figure}

A version tracker is a list where each node records the timestamp of a range query and whether that range query has been completed or not. The structural design of the version tracker can be seen in Figure \ref{fig:version_tracker}. It is essentially a linked list where operations only occur at either end. Whenever a range query starts, its timestamp is added atomically to the tail of the version tracker. Thus the nodes in the linked list are sorted by their timestamp. Please refer to Figure \ref{alg:remHead} for more algorithmic details. 

Whenever a split or merge leaf operation begins, we fetch the timestamp of the oldest range query using the version tracker. Every key marked deleted with a timestamp lesser than the minimum timestamp fetched from the version tracker will not be included in the new nodes created due to splitting or merging. To fetch the oldest range query’s timestamp, we traverse the list from the head and remove the nodes containing finished range queries. The first unfinished node's timestamp is the minimum timestamp among active range queries. Please refer to Figure \ref{alg:remHead} for more algorithmic details.
\label{sec:version_tracker_app}
\end{document}